\documentclass[aps,superscriptaddress,showpacs,nofootinbib,floatfix,epfs]{revtex4}
\pdfoutput=1
\usepackage{amsfonts,amscd,amsmath, amssymb,graphicx,color}

\def\xm{x_{\scriptscriptstyle -}}
\def\xp{x_{\scriptscriptstyle +}}
\def\rm{r_{\scriptscriptstyle -}}
\def\rp{r_{\scriptscriptstyle +}}

\def\ap{\alpha_{\scriptscriptstyle +}}

\def\la0{\lambda_{\scriptscriptstyle 0}}

\def\qqb{\text{\tiny Q}\bar{\text{\tiny Q}}}
\def\qq{\text{\tiny QQ}}
\def\3q{3\text{\tiny Q}}

\def\qu{\text{\tiny Q}}
\def\aqu{\bar{\text{\tiny Q}}}

\def\oh{\frac{1}{2}}
\def\ep{\text{e}}
\def\g{\mathfrak{g}}
\def\oh{\frac{1}{2}}

\def\s{\mathfrak{s}}
\def\rh{r_{\text{\tiny h}}}

\def\erfi{\text{Erfi}}

\def\h0{h_0}
\def\T0{T_{\text{\tiny 0}}}
\def\con{\text{\tiny con}}

\begin{document}
\title{Color screening masses from string models}
\author{Oleg Andreev}
\affiliation{L.D. Landau Institute for Theoretical Physics, Kosygina 2, 119334 Moscow, Russia}
\affiliation{Arnold Sommerfeld Center for Theoretical Physics, LMU-M\"unchen, Theresienstrasse 37, 80333 M\"unchen, Germany}
\begin{abstract} 
We use gauge/string duality to estimate the Debye screening mass at non-zero temperature and baryon chemical potential. We interpret this mass as the smallest one in the open string channel. Comparisons are made with the results from holography and lattice QCD.
 \end{abstract}
\maketitle

\vspace{-7cm}
\begin{flushright}
LMU-ASC 24/16
\end{flushright}
\vspace{6cm}

\vspace{.5cm}
\section{Introduction}
\label{intro}
\renewcommand{\theequation}{1.\arabic{equation}}
\setcounter{equation}{0}

Understanding the properties of quarks and gluons at non-zero temperature and density is of primary importance to both theory and experiment. One of those is the phenomenon of Debye screening which is a common feature of one of the fundamental states of matter - plasma. Its numerical characteristics are screening masses which imply the exponential fall-off of correlation functions at asymptotically large separation. 

There is a long history of computing the screening masses either perturbatively or non-perturbatively.\footnote{See, for example, \cite{review-b} and references therein.} At temperatures of a few times of the deconfinement temperature, which are the temperatures probed by heavy ion collisions, reliable calculations require numerical simulations on the lattice. However, at non-zero baryon density or, equivalently, non-zero baryon chemical potential those are rather limited. Gauge/string duality provides new theoretical tools for studying strongly coupled gauge theories \cite{review-ads} that is a powerful motivation for using it as a possible alternative. Although there still is no string theory dual to QCD, it would seem very good to understand to what extent effective string models already at our disposal are accurate in their ability to mimic or even predict the behavior of the screening masses at non-zero temperature and baryon chemical potential. 

A suitable class of gauge invariant correlators includes the correlators of the Polyakov loops 

\begin{equation}\label{C-cor}
C_{n,\bar n}(\mathbf{x}_1,\dots,\mathbf{x}_n;\bar{\mathbf{x}}_1,\dots,\bar{\mathbf{x}}_{\bar n})=\langle\,
L(\mathbf{x}_1)\dots L(\mathbf{x}_n)L^\dagger(\bar{\mathbf{x}}_1)\dots L^\dagger(\bar{\mathbf{x}}_{\bar n})
\,\rangle
\,,
\end{equation}
with $L=\frac{1}{N}\text{trPexp}\Bigl[ig_{\text{\tiny YM}}\int_0^{1/T}dt\,\text{A}_t\Bigr]$. Here the trace is over the fundamental representation of $SU(N)$, $t$ is a periodic variable of period $1/T$, with $T$ the temperature, $\text{A}_t$ is a time component of a gauge field, and $g_{\text{\tiny YM}}$ is a gauge coupling constant. These correlators are of special interest since they determine the free energy of a configuration of $n$ heavy quarks and $\bar n$ heavy antiquarks placed at positions $\mathbf{x}_1,\dots \mathbf{x}_n$ and $\bar{\mathbf{x}}_1,\dots\bar{\mathbf{x}}_{\bar n}$, respectively \cite{mcl}. 

The present paper continues a series of studies on effective five-dimensional string theories started in \cite{az1,baryons1}. It began with the following question. Given the prescription for computing the correlator $C_{1,\bar 1}$ \cite{theisen}, what is that for $C_{2,0}$? Or, in other words, how to compute a diquark free energy via gauge/string duality? In Sec. III, we will attempt to give such a prescription. In doing so, the notion of a baryon vertex \cite{witten-bv} plays a pivotal role. To cross check the prescription we make two estimates of the Debye screening mass: the first estimate from $C_{1,\bar 1}$ and the second from $C_{2,0}$. We go on in Sec. IV to compare our estimates with those based on holography (gravity modes) and numerical simulations. Finally, we conclude in Sec.V with a discussion of some open problems. In particular, we address the issue of a three-loop correlator $C_{3,0}$. Some technical details are included in the Appendices.

\section{Preliminaries}
\renewcommand{\theequation}{2.\arabic{equation}}
\setcounter{equation}{0}

For orientation, we begin by setting the framework and recalling some preliminary results. We will compute the loop correlators in a 5-dimensional effective string model which is an extension of the model \cite{az1,baryons1} to finite temperature and baryon chemical potential. Now, a key point is that the background geometry is described by a charged black hole. This is a standard approach to describe the deconfined phase in AdS/CFT which is a correspondence between conformal field theory and string theory on Anti de Sitter (AdS) space \cite{review-ads}. The phenomenon of string breaking is modeled by the black hole horizon where the effective string tension vanishes. 

The strings in question are elementary Nambu-Goto strings living in a five-dimensional curved space. The endpoints of each string are electrically charged with respect to a background $U(1)$ field, with charges that are $+1/3$ or $-1/3$, such that the net charge is zero.\footnote{The normalization has been chosen so that a baryon has a charge of $+1$.} Our ansatz for the background fields is 
that

\begin{equation}\label{metric5}
ds^2=w(r){\cal R}^2\Bigl(f(r)dt^2+d\vec x^2+f^{-1}(r)dr^2\Bigr)
\,,
\qquad
A=\bigl(A_t(r),0,\dots,0\,\bigr)\,,
\qquad
\text{with}
\qquad
w(r)=\frac{\ep^{\s r^2}}{r^2}
\,.
\end{equation}
It is a one-parameter deformation of the Reissner-Nordstr\"om solution in Euclidean $\text{AdS}_5$ \cite{chamblin}, with $\s$ a deformation parameter. For $f=1$ it reduces to that of \cite{az1,baryons1} which reproduces the area and $Y$-laws for Wilson loops in the infrared regime. The boundary of this space is at $r=0$. $f$ and $A_t$ are some functions\footnote{We discuss explicit formulas for $f$ and $A_t$ in section IV and the 
Appendix C. } of $r$ subject to the conditions

\begin{equation}\label{G-boundary}
f(0)=1\,,
\qquad
f(\rh)=0
\,,
\end{equation}
and accordingly
\begin{equation}\label{A-boundary}
A_t(0)=\mu\,,
\qquad
A_t(\rh)=0
\,.
\end{equation}
We assume that $f(r)$ decreases with $r$ and has a positive root at $r = \rh$ such that a regular black hole horizon occurs. As usual, $\mu$ is identified with the baryon chemical potential of a dual gauge theory and an inverse period of $t$ with its temperature $T$ \cite{review-ads}.

The simplest correlators, with $n=\bar n=1$ and $n=1$, were computed in \cite{az3,col1} and \cite{a-pol}. A question to be asked is whether the correlators for other $n$ can be computed. At this point, it is good to remember the notion of a baryon vertex which is a higher-dimensional analog of a string junction. In ten dimensions, the baryon vertex is a wrapped five-brane whose world-volume is a product of some internal space $X$ and a time curve in $\text{AdS}_5$ \cite{witten-bv}. Viewed from a five-dimensional perspective, it looks like a point-like object whose action is \cite{baryonsN}

\begin{equation}\label{vertex}
S_{\text{vert}}=\mathfrak{m}\int d\tau\sqrt{f}\,\frac{\ep^{-2\s r^2}}{r}\,.
\end{equation}
The form of the integrand follows from the world-volume term in the five-brane action but $\mathfrak{m}$ being a result of resummation of infinitely many terms ($\alpha'$-corrections) is a parameter of the model. This ansatz is quite successful because it allows us to describe the lattice results for the expectation value of a baryonic Wilson loop by using one free parameter which is $\mathfrak{m}$ \cite{baryonsN}.

In the presence of an external gauge field, we extend the action by adding a coupling to $A_t$:

\begin{equation}\label{Wilson-vertex}
S_{\text{vert}}\rightarrow S_{\text{vert}}+\int d\tau A_t\frac{dt}{d\tau}
\,.
\end{equation}
The motivation for such a form of the coupling is drawn from the AdS/CFT construction \cite{witten-bv}, where the five-brane action has a coupling $\int A\wedge G_5$. Here $G_5$ is a five-form field with $3$ units of flux on $X$. The generalization to an antibaryon vertex is straightforward. It is done by changing the sign of the flux: $3\rightarrow -3$ and, as a result, the sign of the last term in \eqref{Wilson-vertex}.

To compute the free-energy for a time-independent configuration, it is convenient to choose the static 
gauge $t=\tau$. In this case, the action becomes a function of $r$

\begin{equation}\label{v-action}
S_{\text{vert}}=\frac{1}{T}
\Bigl(\mathfrak{m}\sqrt{f}\,\frac{\ep^{-2\s r^2}}{r}+A_t(r)\Bigr)
\,,
\end{equation}
which says how strongly the free-energy depends on the vertex position $r$.

Now we would like to explain how to compute the correlator $C_{n,\bar n}$ in this formalism. First, we place external sources, $n$ quarks and $\bar n$ antiquarks, at the boundary points $\mathbf{x}_i$ and $\bar{\mathbf{x}}_i$ of the five-dimensional space. Next, we consider configurations in which each quark (antiquark) is the endpoint of the Nambu-Goto string. The strings in question are oriented. For each string there is a charge $+1/3$ at one end of the string and a charge $-1/3$ at the other end
such that the net charge is zero. The strings join at baryon and antibaryon vertices in the interior. The total action is therefore the sum of Nambu-Goto and vertex actions together with additional couplings to the background gauge field. The expectation value of $C_{n,\bar n}$ is given by the world-sheet path integral. In practice, it can be evaluated by the saddle-point approximation. The expectation value is then

\begin{equation}\label{C}
C_{n,\bar n}(\mathbf{x}_1,\dots,\mathbf{x}_n;\bar{\mathbf{x}}_1,\dots\bar{\mathbf{x}}_{\bar n})=\sum_{m} w_m \ep^{-S_m}
\,.
\end{equation}
Here the sum goes over all possible string configurations that obey the boundary conditions. $S_m$ represents the minimal action of the $m$-configuration whose weight is $w_m$. This formula is a generalization of that in \cite{malda}. $S_m$ now includes not only the minimal area, but also the contributions from boundaries and vertices. Note that one can, in fact, think of the configurations in \eqref{C} as those of \cite{veneziano} with all but the external sources dipping into the bulk. 

\section{Screening Masses from Correlators of Polyakov Loops} 
\renewcommand{\theequation}{3.\arabic{equation}}
\setcounter{equation}{0}

As an important illustration of these ideas, we will give a few examples with increasing 
complexity. We keep the form of $f$ and $A_t$ completely general and consider the case when the black hole horizon is closer to the boundary than the soft wall \cite{az2}. This implies that $\rh<\frac{1}{\sqrt{\s}}$ and the corresponding gauge theory is deconfined. In addition, we will assume that at short distances string configurations with a minimum number of baryon/antibaryon vertices are dominant. 

\subsection{Two Oppositely Oriented Loops}

We begin by recalling how the correlator $C_{1,\bar 1}$ can be evaluated within the approximation \eqref{C}.  The analysis that follows is applicable for any $N$. For more explanations, see \cite{az3,col1}.

\begin{figure}[htbp]
\centering
\includegraphics[width=7.75cm]{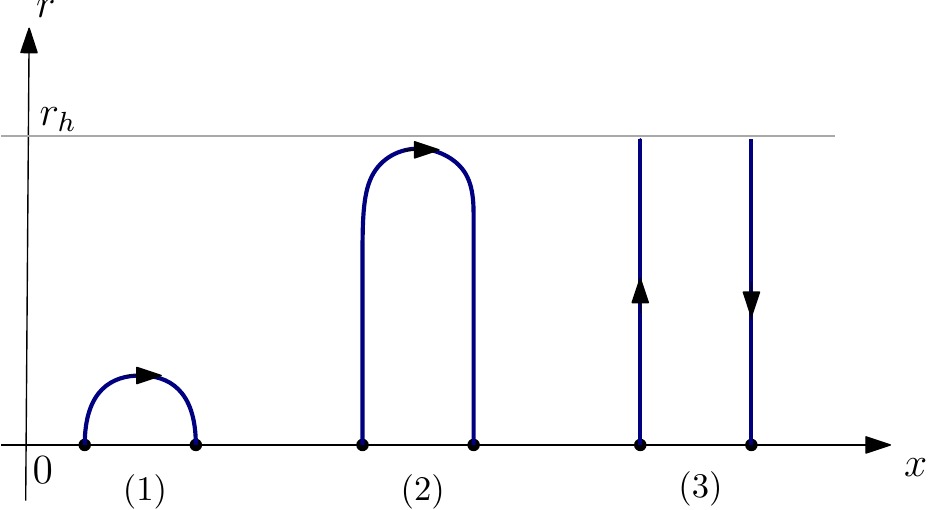}
\caption{{\small String configurations contributing to $C_{1,\bar 1}$. The arrows show the string orientation chosen to go from positive to negative charges. The horizontal line at $r=\rh$ represents the horizon.}}
\label{conf1-1}
\end{figure}

Consider the configurations sketched in Figure \ref{conf1-1}. For convenience, 
we place the sources on the $x$-axis at positions separated by $l$. It is straightforward to write a relation between $l$ and the maximum value of $r$ which holds in the case of the connected configurations. By using the general formula \eqref{l} in the current context, $\rm=0$ and $\ap=0$, we get

\begin{equation}\label{l-con}
l(\nu)=2\rh\nu\int^{1}_0 \frac{dv}{\sqrt{f(\rh\nu\,v)}}
\bigl(\eta^{-1}-1\bigr)^{-\frac{1}{2}}
\,,
\qquad
\eta=\frac{f(\rh\nu)}{f(\rh\nu\,v)}\,v^4\exp\Bigl(2h\nu^2 \bigl(1-v^2\bigr)\Bigr)
\,,
\end{equation}
with $l=\vert x_1-x_2\vert$, $\nu=\tfrac{\rp}{\rh}$, and $h=\s\rh^2$. The parameter $\nu$ takes values in the interval $[0,1]$.\footnote{So $l$ is real for $h<1$.} The extra factor of $2$ is due to the reflectional symmetry (considered modulo string orientation) of the configurations. Note that the function $l(\nu)$ has a local maximum at $\nu=\nu_\ast$ and vanishes at the endpoints of the interval.

Given a solution (string profile), one can evaluate the corresponding action. For the connected configurations, it can be expressed in terms of integrals by using equation \eqref{S0}. Explicitly

\begin{equation}\label{S-con}
S(\nu)=
\frac{2\g}{\nu\rh T}\int^1_{0} \frac{dv}{v^2}\,\Bigl[\ep^{h\nu^2 v^2}
\bigl(1-\eta\bigr)^{-\frac{1}{2}}
-1-v^2
\Bigr]+2c
\,.
\end{equation}
Since the string is neutral, the dependence on the $U(1)$ gauge field drops out. The formulas \eqref{l-con} and \eqref{S-con} are applicable for both cases sketched in Figure \ref{conf1-1}. In the first case, where the string profile is close to the boundary, $\nu$ takes values in the interval $[0,\nu_\ast]$, while in the second, where the string profile looks like a spike, $\nu$ takes values in the interval $[\nu_\ast,1]$. Note that the solutions coincide at $\nu=\nu_\ast$.

For the disconnected configuration, the action reads 

\begin{equation}\label{S-dis}
S=-2\frac{F_0}{T}+2c
\,,
\end{equation}
as it follows from \eqref{S-pol}. $S$ depends only on the minimal area. The reason for this is that the loops are oppositely oriented and, as a result, the two boundary terms cancel each other. 

It is useful for what follows to recall some facts about the string configurations of Figure \ref{conf1-1}. The first is that configurations $(1)$ and $(2)$ exist only if $l$ does not exceed a certain critical value, which is equal to $\rh$ times some numerical factor.\footnote{At zero chemical potential and high temperature, $\rh \sim 1/T$ such that $l T\lesssim 1$ \cite{djg}.} The reason for this is that an effective string tension vanishes at $r=\rh$ so that a string breaks once it touches the horizon. In other words, the phenomenon of string breaking is modeled by a black hole. The second is that configuration $(1)$ dominates at short distances. In that case, $S^{(1)}\sim -1/l$, while the two remaining $S$'s are regular at $l=0$. Third, at long distances the disconnected configuration $(3)$ dominates as in QCD.

Because of scheme ambiguities, the $S$'s may look not so good for a comparison to what is known in the literature. A solution can be found provided the $c$ dependence vanishes from the correlator, and this requires a normalization of $C_{1,\bar 1}$. The normalized correlator is given by 

\begin{equation}\label{C-nor}
\mathbf{C}_{1,\bar 1}(l)=\frac{\langle\, L^\dagger(x_1) L(x_2)\,\rangle}{\vert\langle\, L\,\rangle\vert^2}
\,.
\end{equation}
This correlator does not suffer from linear divergences and, according to \cite{mcl}, determines a difference between the free energies: $\Delta F_{\qqb}=F_{\qqb}-F_{\qu}-F_{\aqu}$ which is usually called the binding free energy of a pair.

Now we want to compute $\Delta F_{\qqb}$ at short distances. For this, we use \eqref{C} together with \eqref{S-con} and \eqref{S-dis}. The resulting formula is 

\begin{equation}\label{DF1-1}
\Delta F_{\qqb}(\nu)=\frac{2\g}{\rh\nu}\int^1_{0} \frac{dv}{v^2}\,
\Bigl[\ep^{h\nu^2 v^2}
\bigl(1-\eta\bigr)^{-\frac{1}{2}}-1-v^2\Bigr]
+
2F_0-T\ln w^{(1)}_{1,\bar 1}
\,,
\end{equation}
where $w^{(1)}_{1,\bar 1}$ is a weight of configuration $(1)$. Since the configuration needed for a description of $C_{1,0}$ or, equivalently, $C_{0,\bar 1}$ is single, we set its weight to $1$. In this form, it is easy to take the limit $\nu\rightarrow 0$ and get rid of $\nu$ with the help of \eqref{l-con}. A short calculation shows that 

\begin{equation}\label{DF1-1l}
\Delta F_{\qqb}(l)=-\frac{\alpha_{\qqb}}{l}+2F_0-T\ln w^{(1)}_{1,\bar 1} +o(l)
\,,
\end{equation}
with $\alpha_{\qqb}=\g (2\pi)^3 \Gamma^{-4}\bigl(\tfrac{1}{4}\bigr)$, which is the same as in \cite{malda}.

How would one determine the exponential fall-off of correlators at long distances? What we must do is to revise the formula \eqref{C}. The most challenging task on this way is of course to find a string dual (if any) to QCD. We have nothing to say here about it and that is why we are dealing with the effective string models. 

One approach to seeing that the series \eqref{DF1-1l} might be useful for gaining insight into the Debye mass involves the use of knowledge obtained from numerical simulations \cite{review-b}. For our purposes, what we need to know can be summarized as follows. The quark-antiquark pair can be in either a color singlet or a color octet state. The free energy of the pair gets the contributions from both states. At short distances, the free energy is dominated by that of the singlet state. One of the parameterizations of the singlet free energy, motivated by high temperature perturbation theory, is given by\footnote{See, e.g., \cite{lattice-karsch,lattice-zantow}.}

\begin{equation}\label{F1}
\Delta F^1_{\qqb}(l)=-\frac{4\alpha}{3 l}\ep^{-ml}
\,,
\end{equation}
with $\alpha$ and $m$ parameters. Like in lattice gauge theory \cite{lattice-karsch,lattice-zantow}, we call $m$ the Debye mass.

Going back to our problem, one important conclusion which emerges is that configuration $(1)$ should correspond to the singlet state.\footnote{Note that this provides us a gauge invariant way to determine the singlet free energy.} It is perfectly possible that at long distances such a classical solution does not exist within effective string models, but it does exist in a full string theory of QCD. We would like to go a step further, however, and assume that at short distances the singlet free energy can be well approximated by the expression \eqref{DF1-1l}. With the help of \eqref{F1}, we obtain an estimate for the Debye mass

\begin{equation}\label{m-g}
m=\frac{1}{\alpha_{\qqb}}\Bigl(2F_0-T\ln w^{(1)}_{1,\bar 1}\Bigr)
\,,
\end{equation}
which is scheme independent. Moreover, the formula is applicable to all models with the asymptotic behavior \eqref{DF1-1l}. 

In the case of interest, we write the Debye mass in a more detailed way

\begin{equation}\label{m}
m=\frac{1}{4\pi^3}\Gamma^{4}\bigl(\tfrac{1}{4}\bigr)\sqrt{\s}
\biggl(
\frac{\ep^h}{\sqrt{h}}-\sqrt{\pi}\,\erfi\bigl(\sqrt{h}\,\bigr)+\mathfrak{w}\frac{T}{\sqrt{\s}}
\biggr)
\,,
\end{equation}
where $\mathfrak{w}=-\ln w^{(1)}_{1,\bar 1}/2\g$. For our purposes, we can think of $m$ as a composite function of $T$ and $\mu$, and treat $\s$ and $\mathfrak{w}$ as the model parameters. In the present section, we try to keep the discussion as general as possible, deferring more detail on $h(T,\mu)$ to Sec.IV. 

We conclude this section with a couple of remarks. First, a configuration which represents a straight string stretched between the boundary and the horizon determines the expectation value of the Polyakov loop. Assuming that it is real, one has 

\begin{equation}\label{L}
\langle\, L\,\rangle =\exp\Bigl\{\frac{1}{T}\Bigl(F_0-\mu_q\Bigr)-c\Bigr\}
\,.
\end{equation}

Second, if we replace $F_0$ by $\langle L\rangle$ in \eqref{m-g}, then it takes the form 

\begin{equation}\label{m-L}
\frac{m}{T}=\frac{2}{\alpha_{\qqb}}
\Bigl(\ln\langle L\rangle+\frac{\mu_q}{T}+c-\oh\ln w^{(1)}_{1,\bar 1}
\Bigr)
\,.
\end{equation}
This provides a relation between the Debye mass and the expectation value of the Polyakov loop. We will return to this relation in Sect. IV.

\subsection{Two Similarly Oriented Loops}

\begin{figure}[htbp]
\centering
\includegraphics[width=7.75cm]{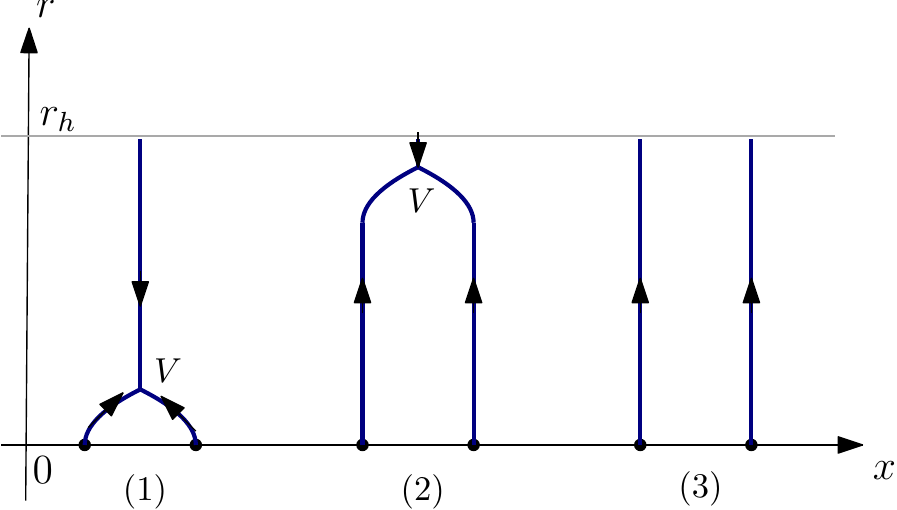}
\caption{{\small String configurations contributing to $C_{2,0}$ at $N=3$. $V$ is a baryon vertex.}}
\label{conf2}
\end{figure}

By considering a string stretched between two sources on the boundary, one gets the correlator $C_{1,\bar 1}$. What does one do with $C_{2,0}$? To obtain it, one can instead consider $N$ strings 
ending on $N$ quark sources and joining at a baryon vertex in the bulk. That, in other words, is a baryonic configuration. The desired configuration is obtained by sending $N-2$ quarks to infinity to decouple them from the remaining two.\footnote{This is perfectly possible in the deconfined phase.} For $N=3$, one is led to suspect that the correlator $C_{2,0}$ can be described in terms of configurations sketched in Figure \ref{conf2}.
\vspace{.5cm}

\vspace{.5cm}
After specializing to $N=3$, we proceed as before. First we evaluate the correlator $C_{2,0}$ by means of equation \eqref{C}. The main additional step required is an analysis of gluing conditions at a baryon vertex which is presented in the Appendix B. Having done this, we then use the result to estimate the Debye mass. The goal is to cross check the estimate obtained in the previous section.

Since the connected configurations are symmetric under reflection through the middle point, the side strings have an identical profile, and the middle string is stretched in the radial direction. Given this, we can write a relation between $l$ and the position of the vertex. Just as in the previous section, we need \eqref{l} at $\rm=0$ but now with $\ap$ determined by the gluing condition \eqref{tan}. Putting all that together gives us

\begin{equation}\label{l20}
l(\nu)=2\rh\nu\int^{1}_0 \frac{dv}{\sqrt{f(\rh\nu\,v)}}
\bigl(\eta^{-1}-1\bigr)^{-\frac{1}{2}}
\,,
\qquad
\eta=\frac{f(\rh\nu)}{f(\rh\nu\,v)}\Bigl(1-\frac{1}{4}\theta^2(\rh\nu)\Bigr)
v^4\exp\Bigl(2h\nu^2 \bigl(1-v^2\bigr)\Bigr)
\,,
\end{equation}
with $l=\vert x_1-x_2\vert$ and $\theta$ defined in \eqref{tan}. The parameter $\nu$ takes values in the interval $[0,\nu_{\text{max}}]$. Here $\nu_{\text{max}}$ is a solution of equation $f(\rh\nu)\bigl(1-\tfrac{1}{4}\theta^2(\rh\nu)\bigr)=0$ such that $0<\nu_{\text{max}}\leq 1$.

\begin{figure}[htbp]
\centering
\includegraphics[width=7cm]{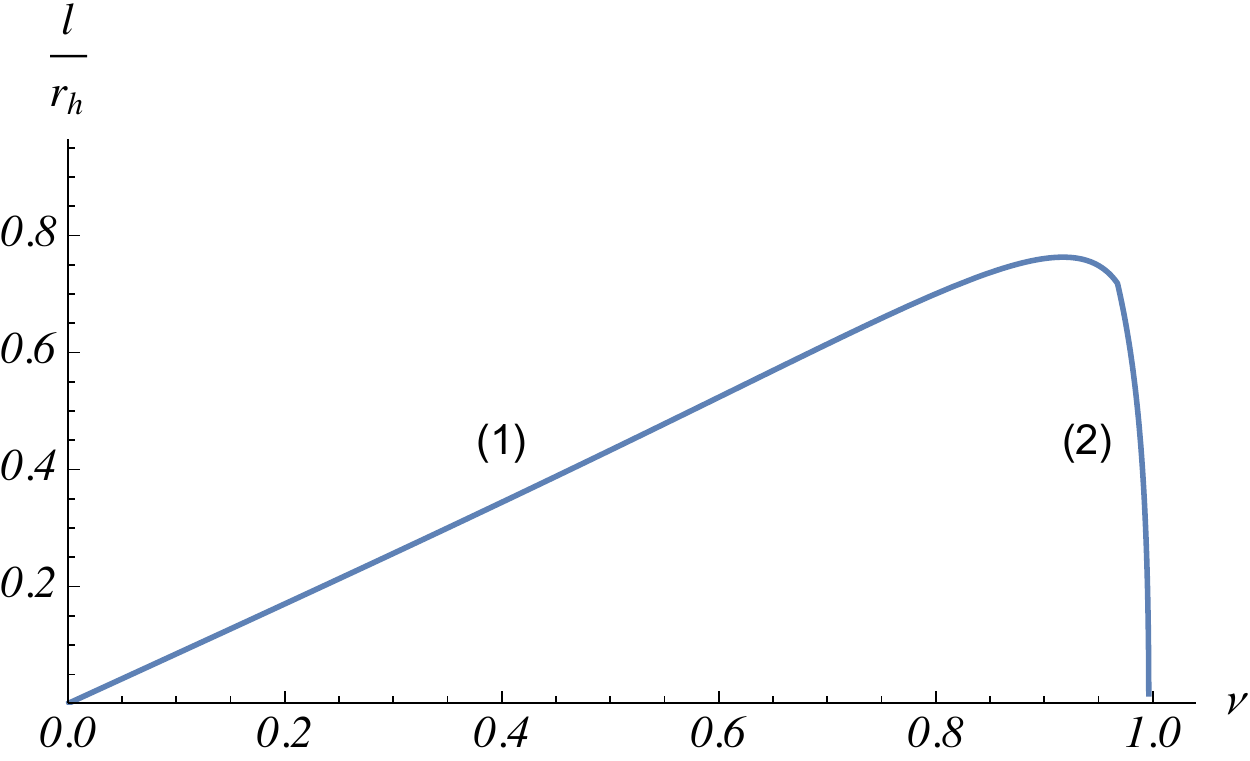}
\caption{{\small $l$ as a function of $\nu$ at $\kappa=-0.083$ and $h=\frac{1}{9}$. It has a maximum at $\nu_\ast\approx 0.917$. Here $\nu_{\text{max}}\approx 0.996$.}}
\label{l2-0}
\end{figure}

At this point, one might ask what is common between the $l(\nu)$'s given by \eqref{l-con} and \eqref{l20}? Both functions are continuous and have zeros exactly at the interval endpoints. The zeros at $\nu\not=0$ appear as a consequence of the condition $\eta=0$, which is possible for the geometry in question. Thus, the $l(\nu)$'s (being regular in the intervals) must be bounded from above. This means that the connected configurations exist only if $l$ does not exceed the critical distance. In addition to these general arguments, let us be more specific and consider what is perhaps the best understood example considered in \cite{az3,col1}, namely the case $f(r)=1-(r/\rh)^4$. Figure \ref{l2-0} shows a typical shape of $l(\nu)$. For configuration (1) with the side strings close to the boundary, $\nu$ takes values in the interval $[0,\nu_\ast]$, while for 
\vspace{.5cm}
\vspace{.5cm}
configuration (2) with the side strings deeply in the interior, $\nu$ takes values in the interval $[\nu_\ast ,\nu_{\text max}]$.

Going further, the action of the middle string can be read from \eqref{S-vert} at $\rp=\rh$, and that of the side strings from \eqref{S0}. Combining these with the action for the vertex, we get 

\begin{equation}\label{S20}
S(\nu)=\frac{\g}{\nu\rh T}
\biggl(
2\int^1_{0} \frac{dv}{v^2}\,
\Bigl[\ep^{h\nu^2 v^2}
\bigl(1-\eta\bigr)^{-\frac{1}{2}}
-1-v^2
\Bigr]
-
\nu\sqrt{\pi h}\,\erfi\bigl(\nu\sqrt{h}\bigr)
+
\ep^{h\nu^2}
+
3\kappa\sqrt{f(\rh\nu)}\,\ep^{-2h\nu^2}
\biggr)
+
\frac{1}{T}\Bigl(2\mu_q-F_0\Bigr)
+
2c
\,,
\end{equation}
with $\kappa=\frac{\mathfrak{m}}{3\g}$ and $\mu_q=\frac{\mu}{3}$. In the last step we used \eqref{S-pol}.  Now the dependence on the gauge field does not drop out, because the configuration is charged under $U(1)$. It comes entirely from the string endpoints on the boundary. The conditions \eqref{A-boundary} and \eqref{v-netcharge} guarantee that there is no dependence on $A_t$ coming from the bulk.

Now let us consider the disconnected configuration. In this case, the total action is twice 
the action of the straight string \eqref{S-pol}. Hence we write it as

\begin{equation}\label{S20-dis}
S=\frac{2}{T}
\Bigl(\mu_q-F_0\Bigr)+2c
\,.
\end{equation}

Having derived the corresponding actions, we are now in a position to draw the same two conclusions as before. The first is that the connected configurations exist only if $l$ does not exceed a certain critical distance, which is equal to $\rh$ times some numerical factor. The second is that configuration $(1)$ dominates at short distances. This requires more explanation. In general, the short distance behavior of $S^{(1)}$ is characterized by the parameter $\kappa$. The conclusion we made is true for some values of $\kappa$. Importantly, the value of $\kappa$ fitted to the lattice is one of those values \cite{baryonsN}. If so, then $S^{(1)}\sim-1/l$, while the other $S$'s are regular at $l=0$.

The correlator $C_{2,0}$ is scheme dependent. This is why any comparison to what is known in the literature   would be pointless unless it is done within the same renormalization scheme. This difficulty can be circumvented by introducing a normalized form of $C_{2,0}$ defined by 

\begin{equation}\label{C20-nor}
\mathbf{C}_{2,0}(l)=\frac{\langle\, L (x_1) L(x_2)\,\rangle}{\langle\, L\,\rangle^2}
\,.
\end{equation}
It follows from this definition that $\mathbf{C}_{2,0}$ determines a difference between the free energies: $\Delta F_{\qq}=F_{\qq}-2F_{\qu}$, with $F_{\qq}$ a diquark free energy. This is nothing else but the binding free energy of a diquark. Note that in addition to the dependence on $c$, the explicit dependence on $\mu_q$ also cancels out.

According to the above formulas, $\Delta F_{\qq}$ is given by 

\begin{equation}\label{FS3}
\Delta F_{\qq}(\nu)=\frac{\g}{\rh\nu}
\biggl(
2\int^1_{0} \frac{dv}{v^2}\,\Bigl[\ep^{h\nu^2 v^2}
\bigl(1-\eta\bigr)^{-\frac{1}{2}}
-1-v^2
\Bigr]
-
\nu\sqrt{\pi h}\,\erfi\bigl(\nu\sqrt{h}\bigr)
+
\ep^{h\nu^2}
+3\kappa\sqrt{f(\rh\nu)}\,\ep^{-2h\nu^2}
\biggr)
+F_0
-T\ln w^{(1)}_{2,0}
\,,
\end{equation}
where $w^{(1)}_{2,0}$ is a weight of configuration $(1)$. Keeping in mind that configuration (1) dominates, it is straightforward to find the two leading terms in the expansion of $\Delta F_{\qq}$ in powers of $l$. A simple but somewhat lengthy calculation reveals that 

\begin{equation}\label{F3short}
\Delta F_{\qq}(l)=-\frac{\alpha_{\qq}}{l}+F_0-T\ln w^{(1)}_{2,0} +o(l)
\,,
\end{equation}
with
\begin{equation}\label{coeff2}
\alpha_{\qq}=-\frac{1}{2}\g\,\xi^{-\oh}B\bigl(\xi^2;\tfrac{3}{4},\tfrac{1}{2}\bigr)
\Bigl(1+3\kappa+\frac{1}{2}\xi^{\frac{1}{2}}B\bigl(\xi^2;-\tfrac{1}{4},\tfrac{1}{2}\bigr)\Bigr)
\,,
\qquad
\xi=\frac{\sqrt{3}}{2}\bigl(1-2\kappa-3\kappa^2\bigr)^\oh
\,.
\end{equation}
Here $B(z;a,b)$ is the incomplete beta function. The coefficient $\alpha_{\qq}$ is the same as that in the static quark-quark potential \cite{baryonsN}. Importantly, it is positive for all values of $\kappa$ in a narrow range which is needed to fit the lattice data on the three quark potential. 

We are now in the same situation as we were with the correlator $\mathbf{C}_{1,\bar 1}$. A string dual to QCD is still missing that obstructs our ability to compute the exponential fall-off of correlators at long distances from the first principles. Here again, we follow the approach relying on a knowledge acquired from numerical simulations. We will not go into great detail about it, only what is needed for our purposes. The diquark can be in either a color antitriplet or a color sextet state. The diquark free energy gets the contributions from both states. At short distances, the free energy is dominated by that of the antitriplet state. One of the parameterizations of the antitriplet free energy, motivated by high temperature perturbation theory, is given by\footnote{See, e.g., \cite{hueb,dor}.}

\begin{equation}\label{F3}
\Delta F^{\bar 3}_{\qq}(l)=-\frac{2\alpha}{3 l}\ep^{-ml}
\,,
\end{equation}
with the same parameters as in \eqref{F1}. If so, then $m$ is the Debye mass.

Now we can draw an interesting conclusion about configuration $(1)$. It should correspond to the antitriplet state. Thus, the string construction provides a gauge invariant way to determine the antitriplet free energy. Of course, it should be perfectly possible, not within effective string models, but in a full string theory of QCD. In undertaking to make an estimate of the Debye mass, a stronger assumption has to be made. Here we assume that at short distances the antitriplet free energy can be well approximated by the expression \eqref{F3short}. Then, with the help of \eqref{F3}, we have the following estimate for the  Debye mass

\begin{equation}\label{m'}
m=\frac{1}{\alpha_{\qq}}\Bigl(F_0-T\ln w^{(1)}_{2,0}\Bigr)
\,.
\end{equation}
It is scheme independent and applicable to all effective string models with the asymptotic behavior \eqref{F3short}. 

We close this section with a few comments:

\noindent (i) As a check of consistency, we can compare the two estimates. At zero temperature, nothing else but the ratio between $\alpha_{\qq}$ and $\alpha_{\qqb}$ matters. Of course, its value may be set to $1/2$ by fitting a value of $\kappa$. This gives $\kappa\approx -0.102$ which is rather close to the value of $-0.083$ obtained from fitting the calculated three-quark potential to the lattice data \cite{baryonsN}. Thus the estimates do look plausible, especially for the effective model we are pursuing. At finite temperature, in addition, a relation between the weights is required. It is $\bigl(w^{(1)}_{1,\bar 1}\bigr)^\oh=w^{(1)}_{2,0}$. At the moment we do not know whether this can be demonstrated within our model. We can only appeal to lattice gauge theory, where the relative weight of the singlet state is $\frac{1}{9}$, while that of the antitriplet $\frac{1}{3}$.

\noindent (ii) In the model of interest, the expression \eqref{m'} becomes 

\begin{equation}\label{m20}
m=-2\,\xi^\oh\sqrt{\s}
\frac{\frac{\ep^h}{\sqrt{h}}-\sqrt{\pi}\,\erfi\bigl(\sqrt{h}\bigr)+\mathfrak{w}\frac{T}{\sqrt{\s}}}
{B\bigl(\xi^2;\tfrac{3}{4},\tfrac{1}{2}\bigr)
\bigl(1+3\kappa+\frac{1}{2}\xi^{\frac{1}{2}}B\bigl(\xi^2;-\tfrac{1}{4},\tfrac{1}{2}\bigr)\bigr)}
\,.
\end{equation}
Here $\mathfrak{w}=-\ln w^{(1)}_{2,0}/\g$. Note that $m$ is positive for the values of $\kappa$ we are considering.

\noindent (iii) For the correlator $C_{0,\bar 2}$, $L$ is replaced by $L^\dagger$, that is equivalent to replacing a quark with an antiquark, and the analysis then proceeds in the same way.

\section{Holographic Models and Lattice} 
\label{continuityequation}
\renewcommand{\theequation}{4.\arabic{equation}}
\setcounter{equation}{0}

What we have learned is the short distance behavior of the correlators which, when combined with the knowledge acquired from numerical simulations, allows us to estimate the Debye mass $m$ and, as a result, the exponential fall-off of the singlet and antitriplet free energies at asymptotically large distances. But what we want to know is the exponential fall-off of the correlators. For example, the connected part of $C_{1,\bar1}$ decays for $l\rightarrow \infty$ as

\begin{equation}\label{LL-M}
\mathbf{C}^{\con}_{1,\bar 1}(l)\sim \ep^{-Ml}
\,,
\end{equation}
where $\mathbf{C}^{\con}_{1,\bar 1}=\mathbf{C}_{1,\bar 1}-1$. $M$ is not equal to $m$. The reason for this is that the correlator gets a contribution from the octet state. Thus, $m$ is not exactly what we want, but it is the first step in the right direction. The second is to determine $M$ in terms of $m$. Here we make the second step and then compare with other results in the literature. In contrast to the first step, it can be done in two, at first sight quite different, ways.\footnote{A bridge that would establish the relation between these two is nothing else but a string dual to QCD.} 

We start again with lattice gauge theory. What is a relation (if any) of $m$ with $M$? The answer to this question is surprising. After taking account of the remaining contribution, the octet/sextet one, the correlator shows the exponential fall-off with the screening mass $M$ well approximated by \cite{review-b}

\begin{equation}\label{Mm}
M=2m
\,.
\end{equation}
This is exactly what we need. It is noteworthy that the above relation also holds at small chemical potential \cite{dor-mu}.

\begin{figure}[htbp]
\centering
\includegraphics[width=9.75cm]{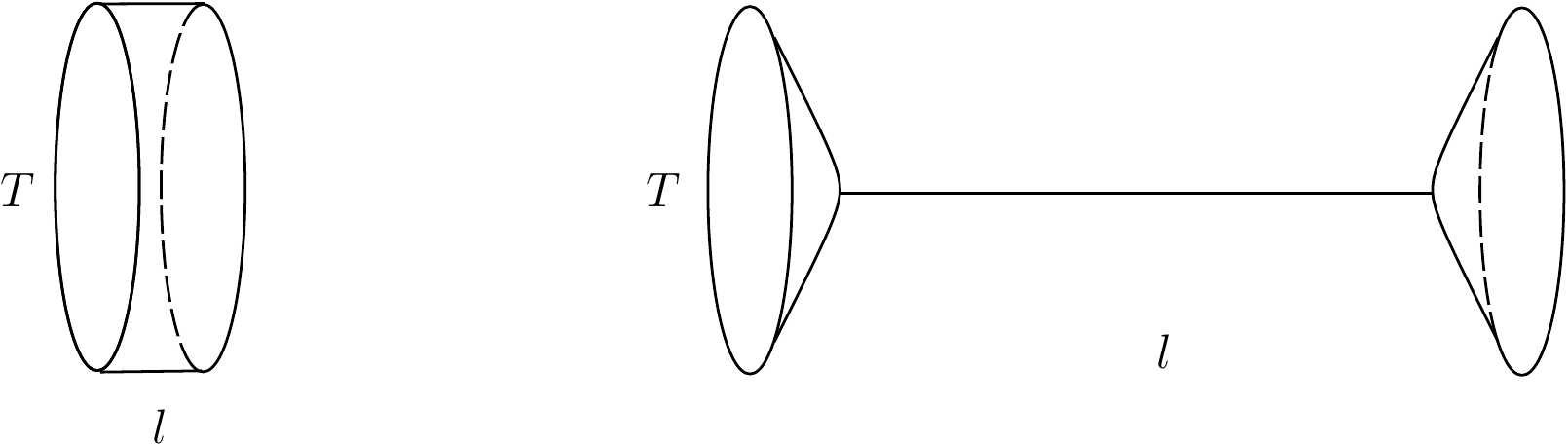}
\caption{{\small Two limiting cases \cite{joe}. Left: The long-strip limit. Right: Closed string states propagating between two boundary states in the long-cylinder limit.}}
\label{cylinder}
\end{figure}

Another way to reach this conclusion is to consider a worldsheet path integral for the correlator $C_{1,\bar 1}$. The worldsheet in question is a cylinder such that each boundary is mapped into a Polyakov loop in a target space. While it is not clear how to properly define and then evaluate this integral for string theory on AdS-like geometries,\footnote{This is the reason why we use the saddle-point approximation \eqref{C}.} in flat space it is well understood \cite{joe}. This is of course not what we need but can be instructive. In the limit $l T\rightarrow 0$ the cylinder looks like a long strip, as shown in Figure \ref{cylinder} on the left, and the leading asymptotics are given by the lightest open 
string states. On the other hand, in the limit $lT\rightarrow\infty$ the cylinder becomes very long, and it looks like that closed string states appear from the vacuum, propagate a distance, and then disappear again into the vacuum, as sketched in Figure \ref{cylinder} on the right. Now the leading asymptotics are given by the lightest closed string states. In this example the relation \eqref{Mm} has a clear meaning: it is a relation between the masses of lightest closed and open string states. 

With this interpretation, we can understand the Debye mass $m$ as a mass of the lightest open string state, while the screening mass $M$ as that of the lightest closed string state. In the present paper, we start with the open string channel and make an estimate for $m$. Next, using the relation \eqref{Mm}, we make it for $M$. This way is a bit longer than one based directly on the lightest (super)gravity modes and proposed in \cite{karch}, but more instructive. Both are two faces of the same coin: strings on AdS-like geometries. 

For future reference, we write $M$ in explicit form

\begin{equation}\label{M}
M=\frac{\gamma}{\pi}\sqrt{\s}
\biggl(\frac{\ep^{h}}{\sqrt{h}}-
\sqrt{\pi}\,\erfi\bigl(\sqrt{h}\,\bigr)+\frac{\mathfrak{w}}{\sqrt{\s}}T
\biggr)
\,,
\qquad
\text{with}
\qquad
\gamma=\frac{1}{2\pi^2}\Gamma^{4}\bigl(\tfrac{1}{4}\bigr)
\,.
\end{equation}
We think of $M=M(h(T,\mu))$ as a composite function. 

\subsection{Sample Models}

So far our discussion was general. We will now give two examples of the function $f(r)$ used in the literature to illustrate the accuracy of the estimate for $M$.

It is natural to begin with the simplest choice of $f$ meaning that the background geometry is the slightly deformed Schwarzschild black hole in $\text{AdS}_5$ space \cite{az2}. Thus, the chemical potential is zero and the function $f(r)$ is 

\begin{equation}\label{fS}
f(r)=1-\Bigl(\frac{r}{\rh}\Bigr)^4
\,.
\end{equation}
The advantage of this choice is that the Hawking temperature is just proportional to the inverse of the horizon position, such that $T=\frac{1}{\pi\rh}$. This allows a great deal of simplification of the resulting equations, and assists in understanding more complicated forms of $f(r)$ later on. For example, in this case, the formula \eqref{M} becomes

\begin{equation}\label{MfS}
\frac{M}{T}=\gamma
\biggl(
\exp\Bigl(\Bigl(\frac{\T0}{T}\Bigr)^2\,\Bigr)-
\sqrt{\pi}\frac{\T0}{T}\,\erfi\Bigl(\frac{\T0}{T}\Bigr)
+\frac{\mathfrak{w}}{\pi}
\biggr)\,,
\end{equation}
where $\T0=\frac{\sqrt{\s}}{\pi}$. In \cite{az2} $\T0$ was interpreted as a critical temperature of the model.

One might think of criticizing the above choice on the grounds that the conditions of Weyl invariance, that is essential to the consistency of string theory \cite{joe}, are not satisfied. In a mathematical language, these are equations which determine all possible backgrounds. In our case the metric is not a solution to equation $\beta^{G}_{\mu\nu}=0$ even to leading order in $\alpha'$. Suppose that the opposite is true. Could one expect that at given $w(r)$ there would exist a corresponding function $f(r)$? In \cite{kiritsis}, it was shown that to leading order in $\alpha'$ this is indeed the case. The two functions are related by a simple differential equation. Such an equation arises as a consequence of the Einstein equations and holds if 
the matter energy-momentum tensor obeys a special constraint. Alternatively, as explained in the Appendix C, one can derive it from the Weyl coefficient for $f$. 

For the case $w(r)=\frac{\ep^{\s r^2}}{r^2}$, the analytical solution to equation \eqref{kiri} with the boundary conditions \eqref{G-boundary} was found in \cite{noro}. It is  

\begin{equation}\label{f-kiri}
f(r)=1-
\frac{1-\bigl(1+\tfrac{3}{2}\s r^2\bigr)\,\ep^{-\tfrac{3}{2}\s r^2}}
{1-\bigl(1+\tfrac{3}{2}h\bigr)\,\ep^{-\tfrac{3}{2}h}}
\,.
\end{equation}
Given this, the corresponding temperature is 

\begin{equation}\label{T-kiri}
T(h)=\frac{9}{8\pi}
\frac{\sqrt{\s}\,h^{\frac{3}{2}}}{\ep^{\tfrac{3}{2}h}-1-\tfrac{3}{2}h}
\,.
\end{equation}
We can think of $T$ as a function of $h$. Thus the screening mass is given in parametric form by $M=M(h)$ and $T=T(h)$. It is worth noting that the function $T(h)$ behaves for $h\rightarrow 0$ as $T=\frac{1}{\pi}\sqrt{\frac{\s}{h}}$. Therefore, this limit is equivalent to the high temperature limit, which turns out to be the same for both choices of $f$. 

The above can be extended to include a background gauge $U(1)$ field. The main additional step required is an analysis of equation $\beta^A_\mu=0$, which is a generalization of Maxwell equation by higher order $\alpha'$-corrections. To leading order in $\alpha'$, it is not technically difficult and presented in the Appendix C. The solutions for $A_t$ and $f$ are given by \eqref{A0} and \eqref{fA}, respectively. Using these solutions, the temperature and baryon chemical potential are 

\begin{equation}\label{tm}
T(h,q)=\frac{9}{8\pi}\frac{\sqrt{\s}(1-q^2)\,h^{\frac{3}{2}}}{\ep^{\frac{3}{2}h}-1-\frac{3}{2}h}
\,,\qquad
\mu(h,q)=\frac{2\sqrt{3\s}\mathfrak{r}\,q \bigl(1-\ep^{-\frac{1}{2}h}\bigr)}{\bigl(9+(7+6h)\ep^{-2h}-16\ep^{-\frac{1}{2}h}\bigr)^{\frac{1}{2}}}
\,.
\end{equation}
Now we can think of $T$ and $\mu$ as functions of $h$ and $q$. The variable $q$ is related to a black hole charge and takes values in the interval $[0,1]$. Note that at $q=0$ the black hole is not charged. 

Finally, the screening mass is given in parametric form by $M=M(h)$ together with $T=T(h,q)$ and $\mu=\mu(h,q)$.

\subsection{Numerics}

Since we have estimated that the screening mass $M$ is about given by \eqref{M}, we can make a comparison with the known results to decide whether it is plausible. 

\begin{figure}[htbp]
\centering
\includegraphics[width=5.75cm]{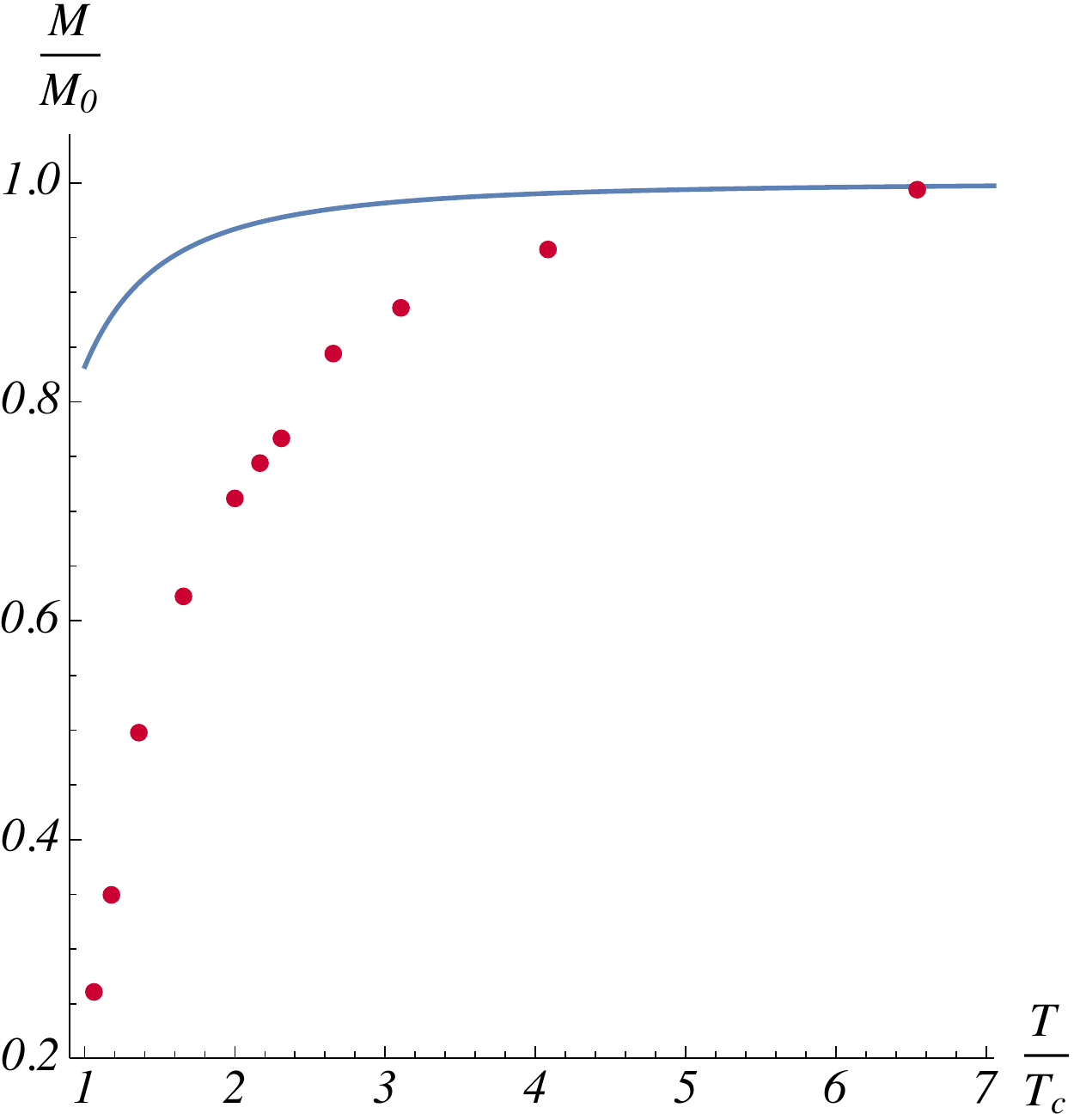}
\hspace{2.75cm}
\includegraphics[width=6.9cm]{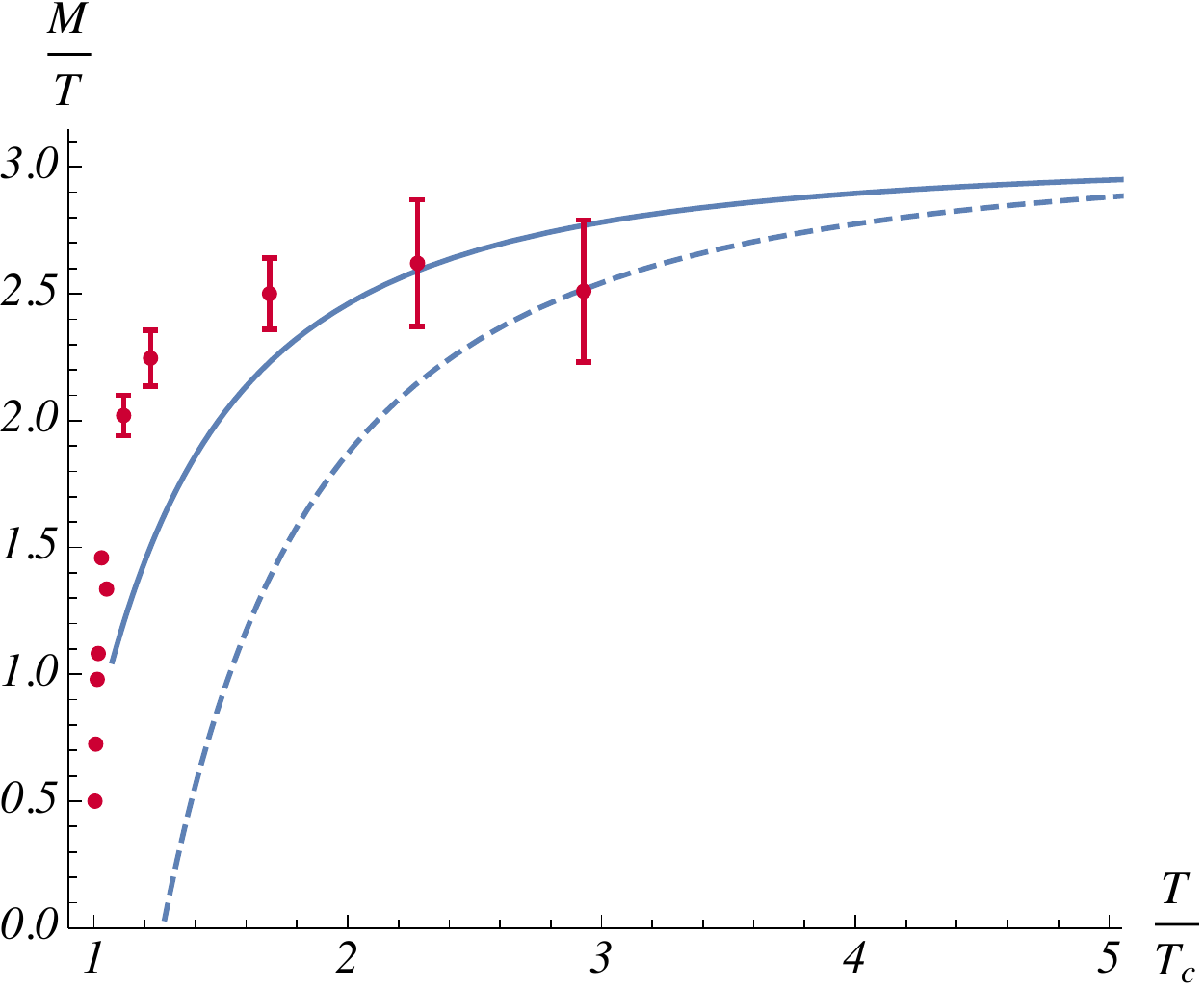}
\caption{{\small Screening mass vs temperature. The solid curve is defined by \eqref{M} and \eqref{T-kiri}. Left: The results of numerical holography \cite{noro}. Here $\mathfrak{w}=0.70$ and $c_4=0.1$ which is the smallest value among all possible ones. Right: The lattice data from \cite{lattice-karsch}. The dashed curve corresponds to \eqref{MfS}. Here we set $\mathfrak{w}=-2.05$.}}
\label{pureglue}
\end{figure}

We begin with a class of holographic models which was studied in \cite{noro}. This seems reasonable because a particular form of the warp factor considered there is similar to ours. In the holographic description the screening mass $M$ comes from the lightest (super)gravity mode which in the present case is an axion. Importantly, the action of the axion field includes an additional warp factor $Z(r)$ parameterized by two parameters $c_0$ and $c_4$. The axion mass is computed numerically for $c_0=1$ and several values of $c_4$. In Figure \ref{pureglue} on the left, we plot the screening mass versus temperature. Units are 
set by $T_c=0.49\sqrt{\s}$ and $M_0=3.40\pi T$, as in \cite{noro}. Since $M_0=M\vert_{\s=0}$, which 
is an asymptotic expression for the limiting behavior of $M$ at large $T$, we set $\mathfrak{w}=0.70$. We see that both approaches result in a similar qualitative behavior: the function $M$ is monotonically increasing with $T$. It has a steep rise near $T=T_c$, but is slowly varying for larger $T$. One might expect that a more quantitative 
agreement requires smaller values of $c_4$ that will lead to a steeper rise near $T=T_c$.

We go on with a $SU(3)$ pure gauge theory. In doing so, we write the critical temperature $T_c=0.629\sqrt{\sigma}$ \cite{boyd} as $T_c=0.435\sqrt{\s}$, where we used 
$\sigma=\ep\g\s$ and in the last step set $\g=0.176$ \cite{baryonsN}. We then plot the results in Figure \ref{pureglue} on the right. It is clear that our estimate is in qualitative agreement with the lattice. There are two important observations. The first observation is that the estimate based on \eqref{T-kiri}, stemming from a more consistent background, agrees better with the lattice. The second is that the lattice results show a much steeper rise in the vicinity of  the critical point. A similar observation was made in relation to the expectation value of the Polyakov loop (correlator $C_{1,0})$ \cite{a-pol}. This might point to the need for quasi-classical corrections in \eqref{C}. In other words, string fluctuations become essential near the critical point.  

\begin{figure}[htbp]
\centering
\includegraphics[width=7.25cm]{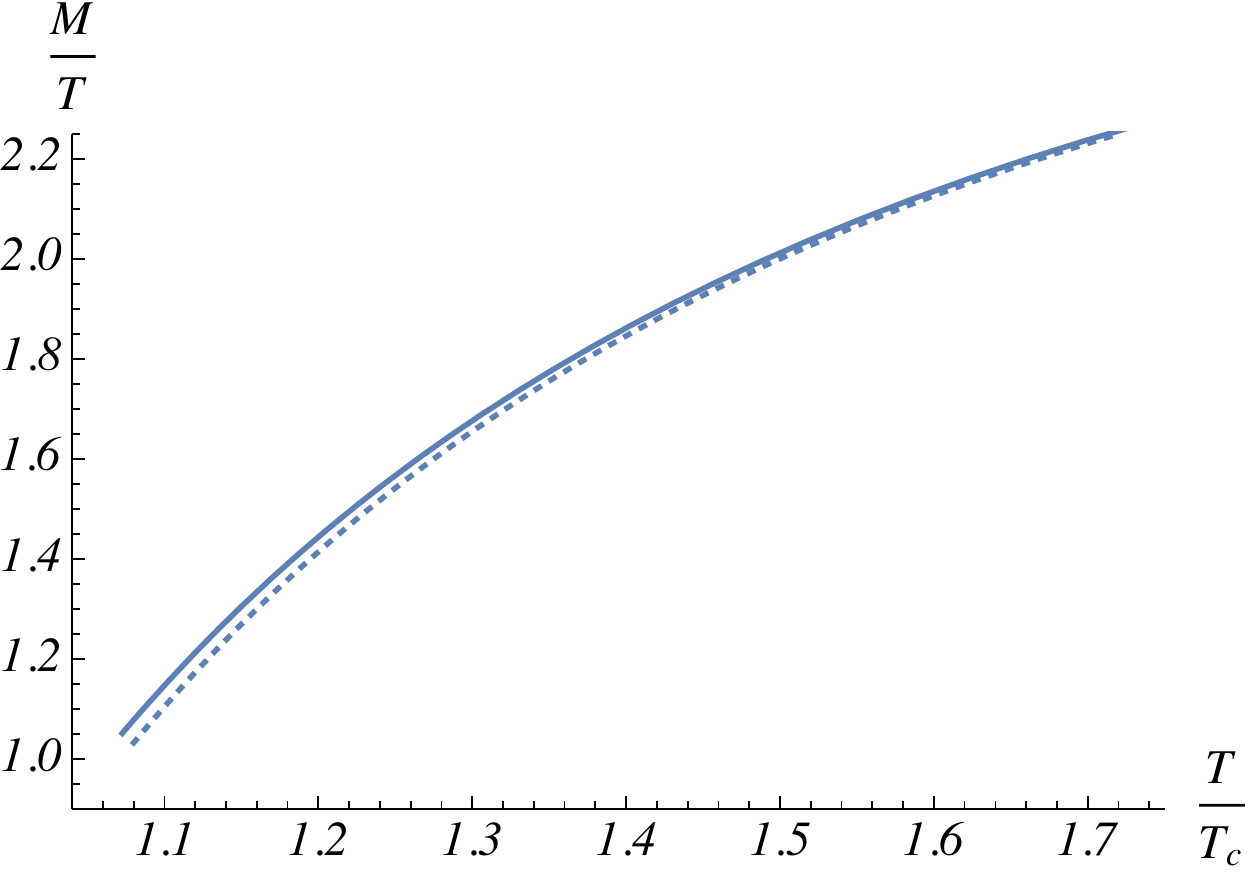}
\hspace{1.8cm}
\includegraphics[width=7.25cm]{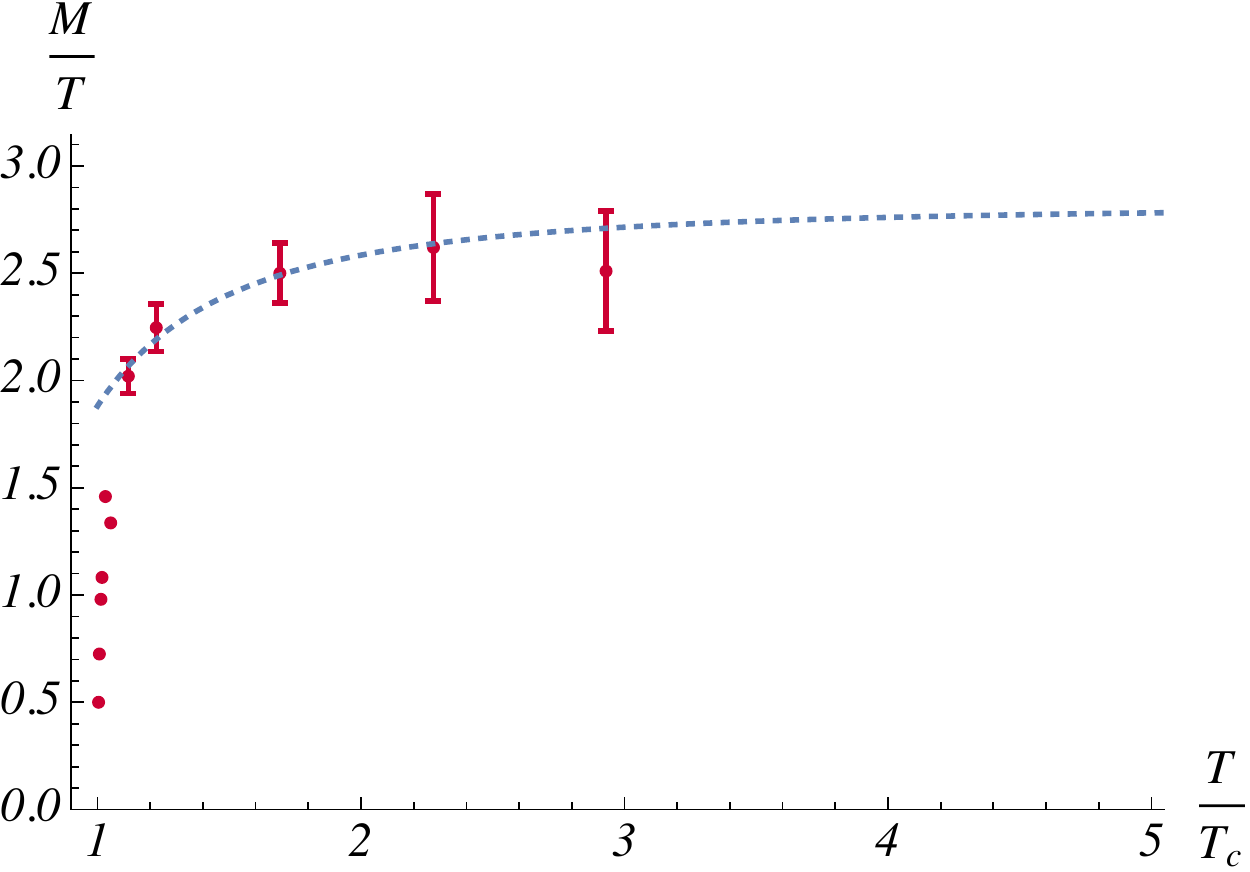}
\caption{{\small A comparison of different $M(T)$ curves for a pure $SU(3)$ gauge 
theory. Left: The solid curve is that of Figure \ref{pureglue}, while the dotted curve corresponds 
to the power law \eqref{Ms0}. Right: The power law at $\mathfrak{w}=-2.13$ and $a=0.107$.}}
\label{pureglue2}
\end{figure}

For practical purposes, the parametric expression for the screening mass $M$ looks somewhat awkward. In \cite{a-pol}, we learned that a reasonable approximation for $C_{1,0}$ ($F_{\qu}$) can be obtained by studying its high-temperature behavior. It is hopefully clear from \eqref{m-L} that a similar derivation for $M$ would proceed in essentially the same way and give a series in powers of $\frac{T_c^2}{T^2}$. So we expand $M(h)$ and $T(h)$  near $h=0$ and then reduce the two equations to a single equivalent equation

\begin{equation}\label{Ms0}
\frac{M}{T}\approxeq\gamma 
\biggl(1+\frac{\mathfrak{w}}{\pi}-a\frac{T_c^2}{T^2}\biggr)
\,,
\end{equation}
where we drop the higher order terms. In this formula $a$ is given by $a=\oh\frac{\s}{\pi^2 T_c^2}$ such that, for the above example, it is $a=\oh (0.435\pi)^{-2}\approx 0.268$. As can be seen from Figure \ref{pureglue2} 
\vspace{0.5cm}
\vspace{0.5cm}
on the left, this is a good approximation. Above $1.3\,T_c$ the discrepancy between the two expressions is negligible. It becomes more and more visible as $T$ approaches $T_c$.

It is tempting to fit the lattice data \cite{lattice-karsch} to the power law \eqref{Ms0}. In Figure \ref{pureglue2} on the right, we have plotted the result. We see that the power law is a good approximation above $1.2\,T_c$, but it fails in the vicinity of the critical temperature. 

In \eqref{Ms0}, the $\frac{1}{T^2}$ term can be determined without any reference at all to concrete models (e.g., those of Section A). Let us explain it in more detail. Going back to the formulas \eqref{m-L} and \eqref{Mm}, we see that the screening mass depends linearly on $\ln\langle L\rangle$. On the other hand, it was suggested in \cite{megi} that the temperature dependence of $L$ is governed by the Gaussian ansatz $\langle L\rangle \sim\ep^{-\frac{C}{T^2}}$. Combining these statements, we find the $\frac{1}{T^2}$ term. In fact,  
we might expect that such a term would be another example of resummation of QCD long perturbative series.\footnote{For a discussion of this issue and its connection to quadratic corrections, see \cite{za}.} This time it is within high temperature perturbation theory.

\begin{figure}[htbp]
\centering
\includegraphics[width=7cm]{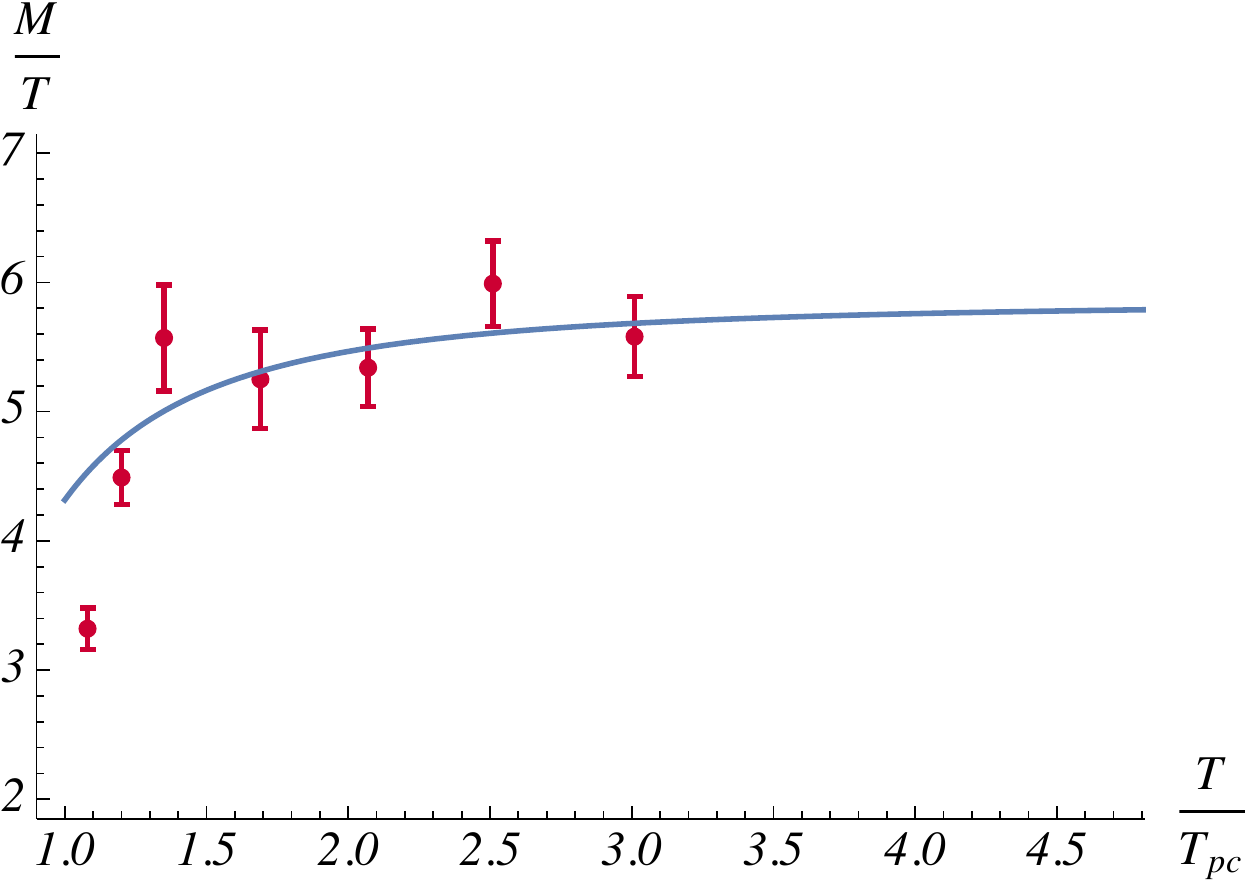}
\hspace{1.8cm}
\includegraphics[width=7cm]{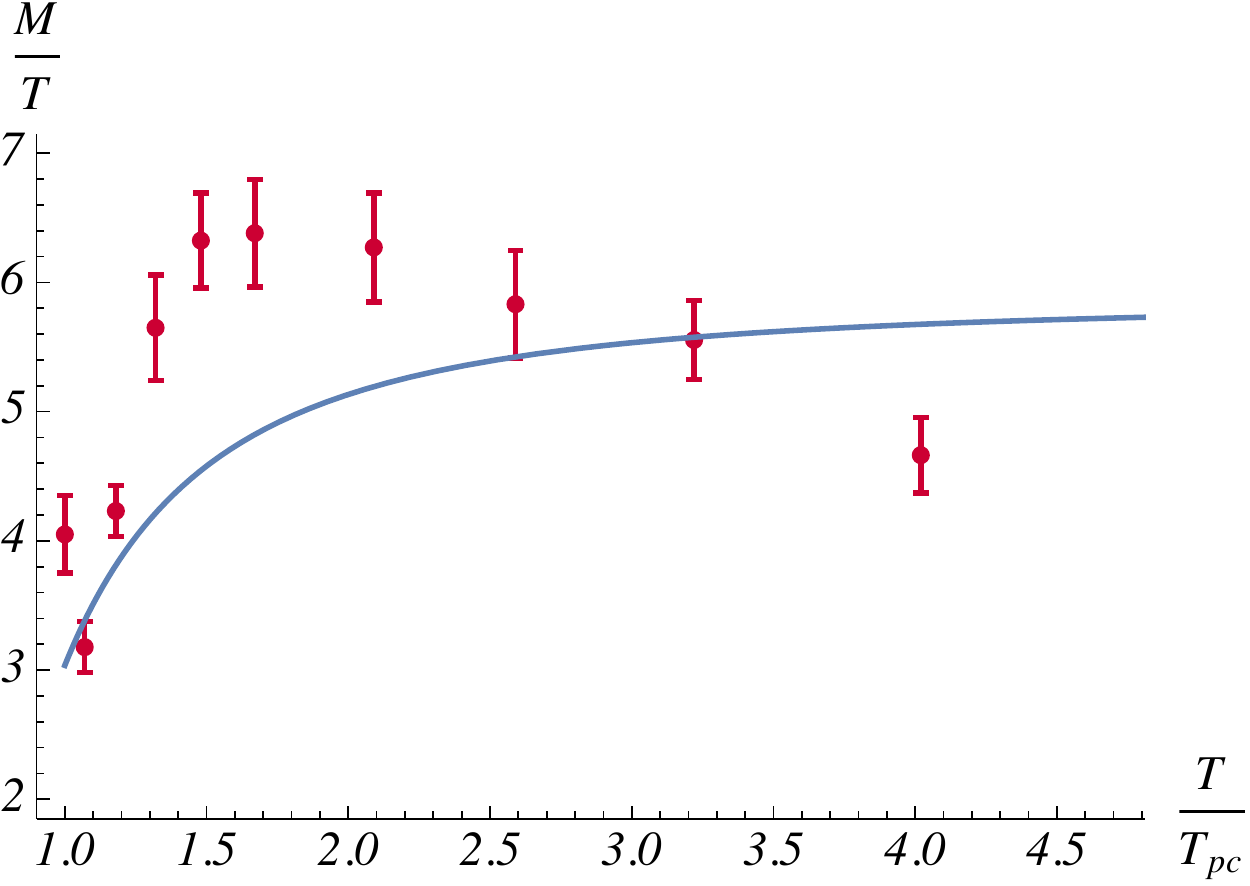}
\caption{{\small Screening mass vs temperature at zero chemical potential and different quark masses \cite{lat-mae}. The solid curve is defined parametrically by \eqref{M} and \eqref{T-kiri}. We set $\mathfrak{w}=-1.04$. Left: Results at $\frac{m_{\text{\tiny PS}}}{m_{\text{\tiny V}}}=0.80$ and $T_{pc}=357\,\text{MeV}$. Right: Results at $\frac{m_{\text{\tiny PS}}}{m_{\text{\tiny V}}}=0.65$ and $T_{pc}=262\,\text{MeV}$.}}
\label{maezawa}
\end{figure}

Now we want to discuss a $SU(3)$ gauge theory with fermions at non-zero baryon chemical potential. We restrict to the case of two flavors. It is motivated by the fact that the singlet free energy is parameterized in the same way in both cases, $N_f=0$ \cite{lattice-karsch} and $N_f=2$ \cite{lattice-zantow}. Our hope is that in the absence of first-principles methods, like lattice QCD, we can gain some important understanding of the theory near the critical line in the $\mu T$-plane. Given the formal formulas that we have just described, it seems straightforward to apply those and determine the screening mass as a function of $T$ and $\mu$. To do it this way however requires a caveat. The models we are considering have no explicit dependence on quark masses. On the other hand, the dependence is visible on the lattice. This can be seen in Figure \ref{maezawa} which shows the results of \cite{lat-mae}. Here the screening mass  was  
\vspace{.5cm}
\vspace{.5cm}
calculated at zero chemical potential and two different values of the ratio between pseudoscalar and vector meson masses at zero temperature.\footnote{Of course, the case of physical quark masses is of primary interest but what happens there is not exactly known.} In contrast to that, the shape of $M(T)$ defined by \eqref{M} and \eqref{T-kiri} slowly depends on $T_{pc}$. Here we set $\s=0.45\,\text{GeV}^2$ as it follows from the $\rho$ meson Regge trajectories \cite{a-q2}. However one common feature does emerge: a rise in the vicinity of the pseudocritical temperature.  

The next case where lattice results are still available is that of small baryon chemical potential. Like in lattice gauge theory \cite{dor-mu}, we expand $M(T,\mu)$ in powers of $\frac{\mu}{T}$

\begin{equation}\label{mu-expansion}
M(T,\mu)=\sum_{n=0}^\infty M_{2n}(T)\Bigl(\frac{\mu}{T}\Bigr)^{2n}
\,,
\end{equation}
for $\frac{\mu}{T}\ll 1$. The Taylor coefficients $M_{2n}$ can directly be derived from \eqref{M} and \eqref{tm}. The first two are given in the Appendix D. Note that when we make this expansion, odd powers of $\frac{\mu}{T}$ do not appear because $h$ is an even function of $\mu$. This is in agreement with the lattice. 

The alternative to what we have described is to expand the Debye mass. In this case, the formula \eqref{Mm} says that $m_{2n}=\oh M_{2n}$. Such a relation was checked numerically \cite{dor-mu} for a few values of $n$. It holds in a wide range of temperatures with a lower bound close to $T_c$. This is a good piece of evidence that the relation \eqref{Mm} is true.

\begin{figure}[htbp]
\centering
\includegraphics[width=7.5cm]{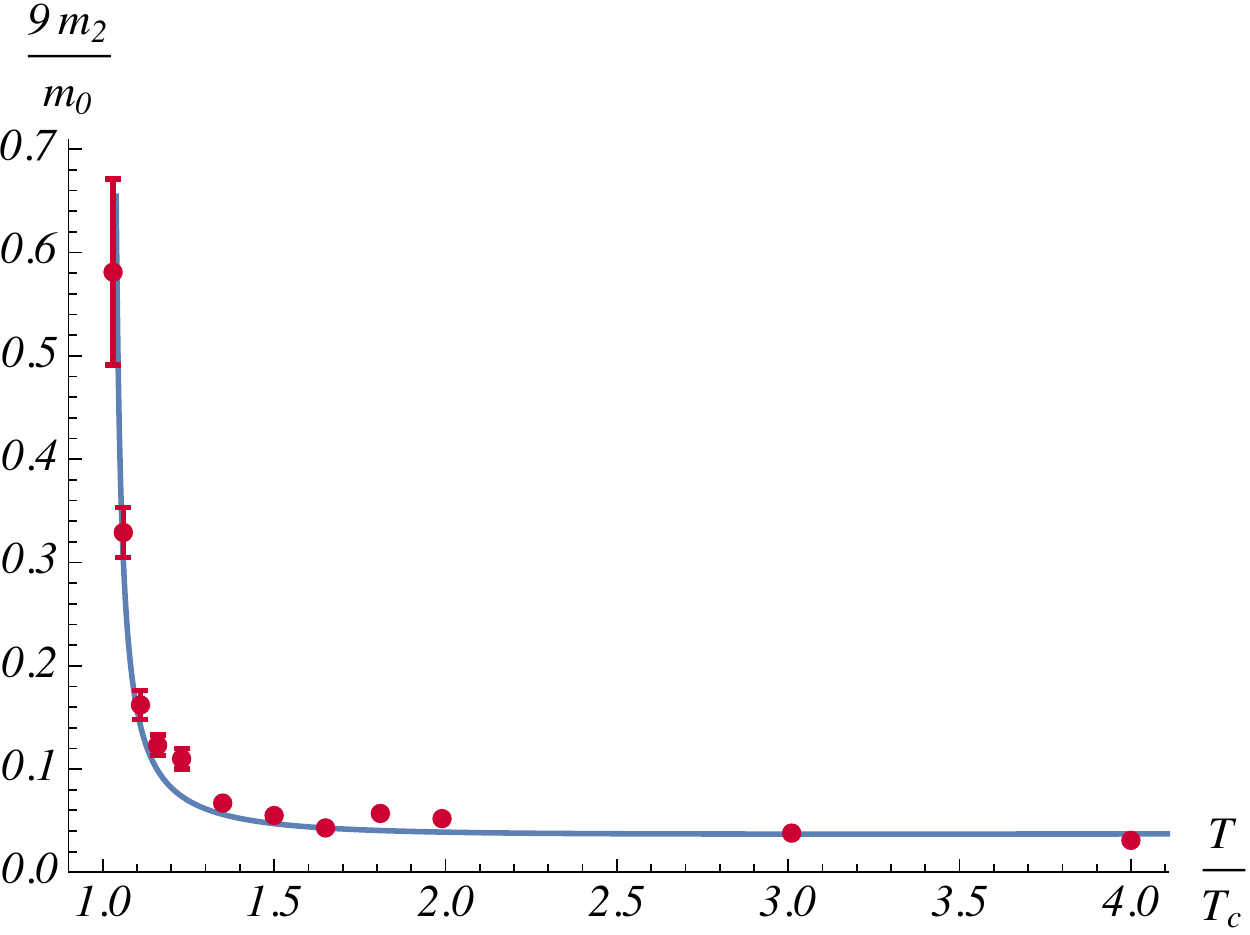}
\hspace{2cm}
\includegraphics[width=6.7cm]{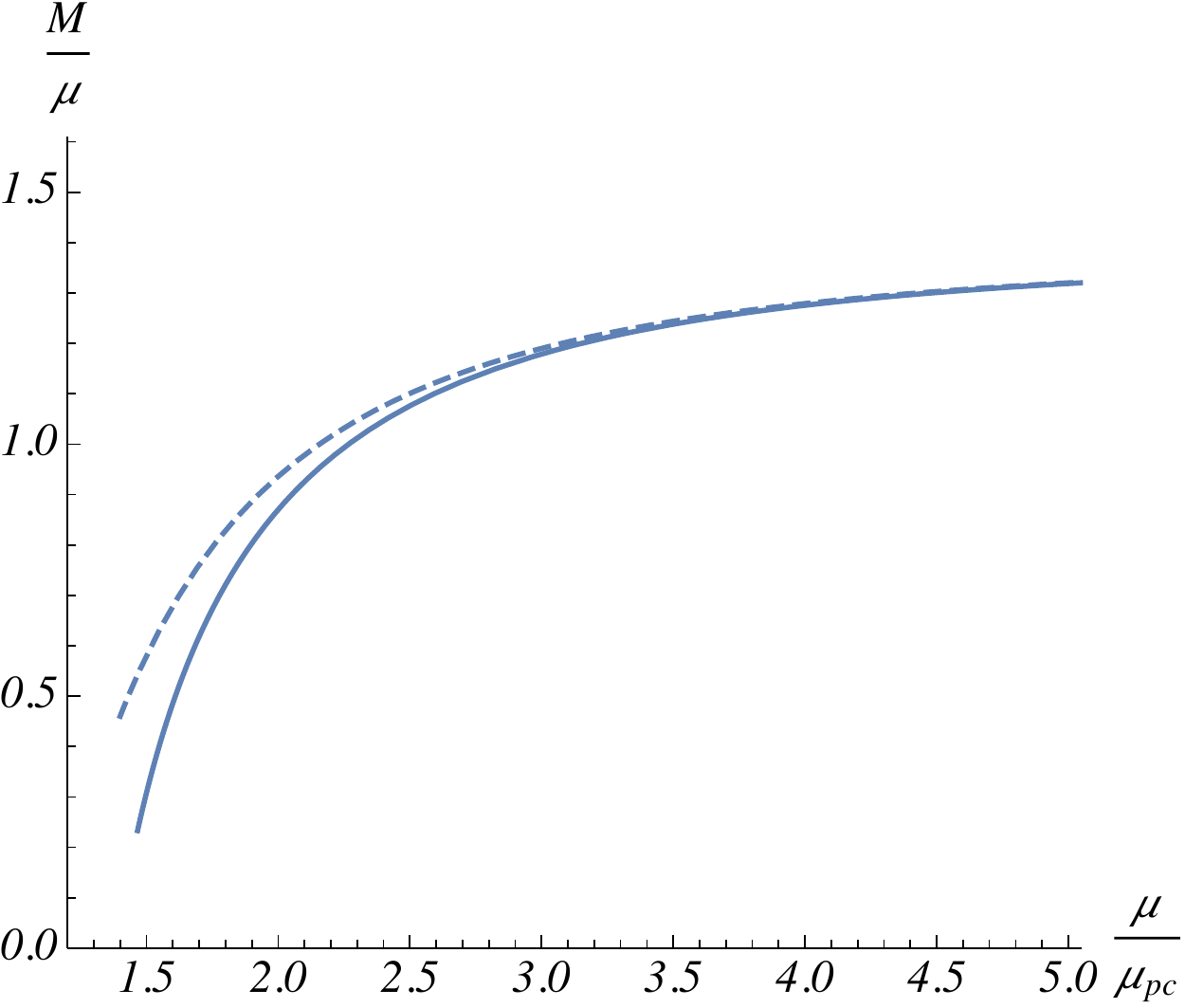}
\caption{{\small Left: Ratio $\frac{9m_2}{m_0}$ versus temperature. The lattice data are taken from \cite{dor-mu}. The curve is defined parametrically by \eqref{T-kiri} and \eqref{M2}. We set $\mathfrak{w}=-1.04$, $\mathfrak{r}=6.00$, and $T_{c}=0.26\sqrt{\s}$. Right: The screening mass versus baryon chemical potential. The solid curve is defined by \eqref{M} and \eqref{tm}, while the dashed curve by \eqref{as-mu}. We set $\mathfrak{r}=2.00$ and $\mu_{pc}=2\sqrt{\s}$.}}
\label{m2}
\end{figure}

Another relation, which was also checked numerically, is that between the coefficients $m_0$ and $m_2$. In \cite{dor-mu}, it was observed that the perturbative result $\frac{m_2}{m_0}=\frac{1}{24\pi^2}$ remains reliable even for $T$ as low as $2T_c$.\footnote{In \cite{dor-mu}, the screening masses are expanded in powers of the quark chemical potential and therefore $m_{2n}^q=3^{2n}m_{2n}$.} Let us confront it with our findings. To do so, we plot the results in Figure \ref{m2}, on the left. In contrast to Figure \ref{maezawa}, 
\vspace{.5cm}
\vspace{.5cm}
agreement with the lattice is very good. Therefore we might expect that the the explicit dependence on quark masses, or the number of flavors, would be multiplicative. 

In view of the good agreement with the lattice, it appears natural to use the perturbative 
result for purposes of determining $\mathfrak{r}$ in terms of $\mathfrak{w}$. Then, from \eqref{M2/M0}, we learn that it is written as

\begin{equation}\label{r}
\mathfrak{r}=2\sqrt{6}\Bigl(1+\frac{\mathfrak{w}}{\pi}\Bigr)^{-\oh}
\,.
\end{equation}
This reduces the number of parameters to just one, which is $\mathfrak{w}$.

One can analyze in a similar fashion the opposite case $\frac{T}{\mu}\ll 1$. It follows from \eqref{tm} that temperature is zero at $q=1$. In that case, $M$ is a function of $\mu$. It is given in parametric form by equations \eqref{M} and \eqref{tm}. In Figure \ref{m2} on the right, we plot the resulting prediction. What emerges is more or less similar to that of Figure \ref{pureglue}: the function $M$ is monotonically increasing with $\mu$. It has a steep rise near $\mu=\mu_{pc}$, but is slowly varying for larger $\mu$. One might expect that such a behavior would indicate that a phase transition occurs near $\mu=\mu_{pc}$.

Actually, as discussed above, one can try to approximate $M(\mu)$ by a power law. To see this, we expand $M(h)$ and $\mu(h)$ near $h=0$ and then reduce these equations to 

\begin{equation}\label{as-mu}
\frac{M}{\mu}\approxeq
\frac{\gamma}{\pi\mathfrak{r}}
\biggl(1-b\frac{\mu^2_{pc}}{\mu^2}\biggr)
\,,
\end{equation}
with $b=\frac{21}{16}\frac{\mathfrak{r}^2\s}{\mu_{pc}^2}$. We drop the higher order terms in the 
expansion. As seen from Figure \ref{m2}, this becomes a good approximation for $\mu$ larger than $2.5\mu_{pc}$. Interestingly, the leading correction is again quadratic. 

Next, we expand the screening mass in powers of $\frac{T}{\mu}$

\begin{equation}\label{T-expansion}
M(T,\mu)=\sum_{n=0}^\infty M_{n}(\mu)\biggl(\frac{T}{\mu}\biggr)^n
\,.
\end{equation}
In contrast to the Taylor series \eqref{mu-expansion}, it contains all integer powers. The coefficients $M_{n}$ can be derived from \eqref{M} and \eqref{tm}. The first two are given in the Appendix D. Note that 
the coefficients of the Taylor series of $m(T,\mu)$ are related to $M_n$ as $m_n(\mu)=\oh M_n(\mu)$. In that case, the plot $\frac{m_1}{m_0}(\mu)$ has a shape similar to that in Figure \ref{m2} on the left so that this ratio also tends to a constant value as $\mu$ goes to infinity.

\begin{figure}[htbp]
\centering
\includegraphics[width=7.75cm]{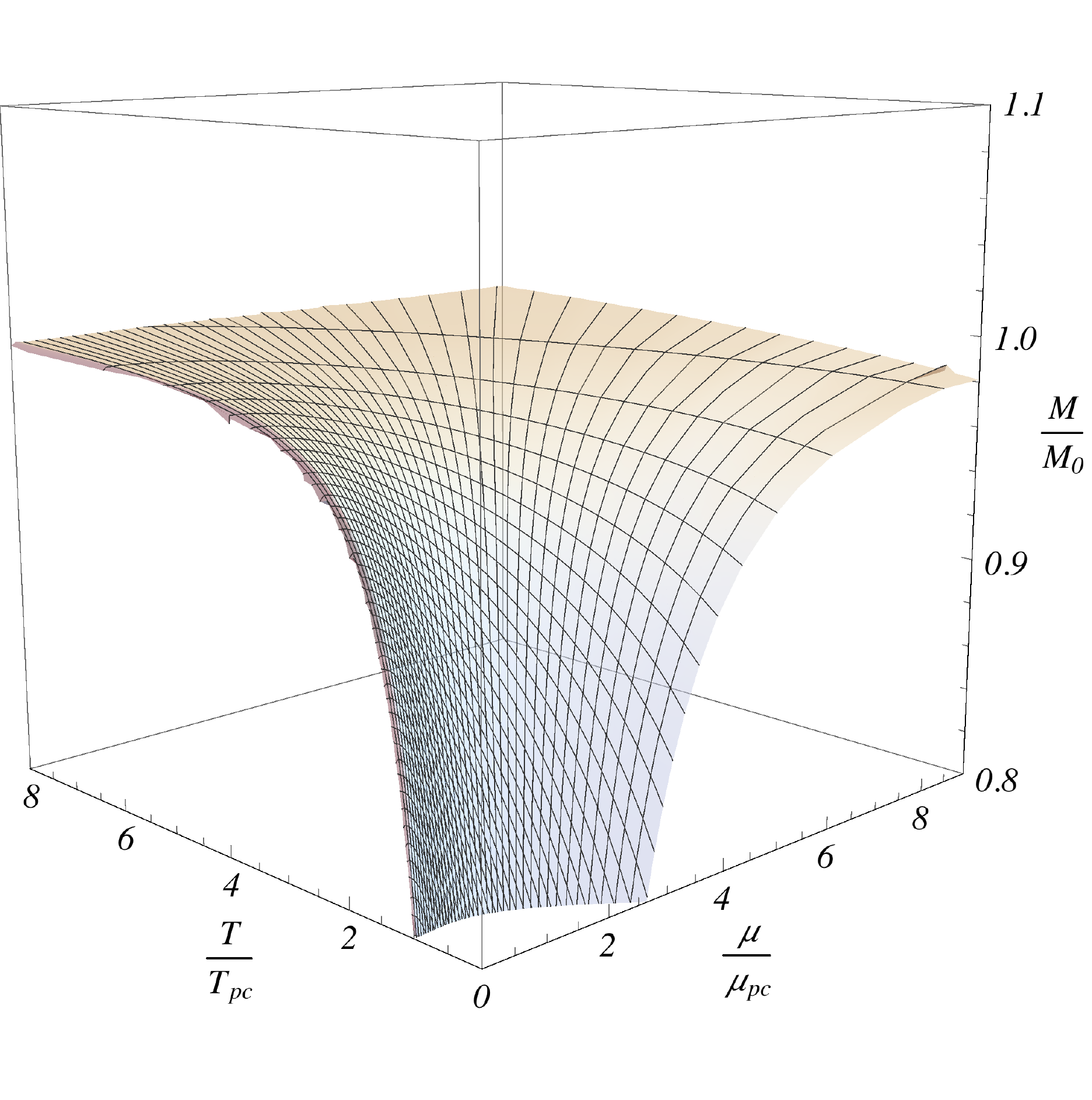}
\caption{{\small $M$ as a function of $T$ and $\mu$. We set $\mathfrak{r}=3.00$,
$T_{pc}=0.30\sqrt{\s}$, and $\mu_{pc}=3\sqrt{\s}$.
}
\\\vspace{.5cm}
}
\label{t-m}
\end{figure}

Finally, there is no difficulty to understand the general case. This means that the screening mass is defined parametrically by equations \eqref{M} and \eqref{tm}, with $h$ and $q$ the parameters. In Figure \ref{t-m}, we plot $M(T,\mu)$. The 
\vspace{.5cm}
normalization is chosen to be $M_0=\frac{\gamma}{\pi}\Bigl( (\frac{\mu}{\mathfrak{r}})^2
\Bigl[\sqrt{(\frac{\mu}{\mathfrak{r}})^2+(\frac{\pi T}{2})^2}
-\frac{\pi T}{2}\Bigr]^{-1}+\mathfrak{w} T\Bigr)$. It is an asymptotic expression for the limiting behavior of $M$ at large $T$ and $\mu$. As before, $M$ shows a steep rise in the region close to the origin and becomes slowly varying for larger values of $T$ and $\mu$. One might expect that such a steep rise would indicate the deconfinement critical line in the $T\mu$-plane. If so, its shape, which looks like that of the bottom edge in Figure \ref{t-m}, seems to be similar to that of \cite{cab}. In the framework of effective (field theory) models, the phase structure of two flavor QCD was studied at different quark masses.\footnote{See, e.g., \cite{tuo} and references therein.} Interestingly, the obtained results show that the Cabibbo-Parisi pattern of the phase diagram holds for a wide range of masses. 

A natural question to ask is, what about $N_f=3$? Obviously, Figure \ref{t-m} shows no sign of crossover. The reason is either it is far from the critical line in the $T\mu$-plane, or the model has nothing to do with it. There are also many other questions to ask. For instance, what can it say about quarkyonic matter? Unfortunately, no real answers will be given here. Our goal is to give an example 
how the screening mass can be estimated in the effective string theory approach and gain some insight that will help us with the further development of the model, as well as with finding a string description of QCD.

\section{Concluding Comments} 
\renewcommand{\theequation}{5.\arabic{equation}}
\setcounter{equation}{0}

(i) It is common practice in phenomenology to use Lipkin's rule which postulates that at zero temperature the quark-quark and quark-antiquark potentials are related by $V_{\qq}=\oh V_{\qqb}$.
Such a simple relation is motivated by the results of perturbative QCD. But one fact makes it difficult to take this seriously: the linear combination $V_{\qq}-\oh V_{\qqb}$ is scheme-dependent. 

There is one important situation in which the problem of scheme-dependence can be avoided. This is the case of binding free energies at finite temperature and non-zero baryon chemical potential. As we saw in Sec.III, in the deconfined phase the binding free energies of two-quark states are scheme-independent. From this point of view the comparison between $\Delta F^1_{\qqb}$ and $\Delta F^{\bar 3}_{\qq}$ may prove to be instructive.  So far, we have exploited the two leading terms in the short distance expansion of the binding free energies. What those suggest is a Lipkin-like relation $\Delta F^{\bar 3}_{\qq}=\oh \Delta F^{1}_{\qqb}$. 
\begin{figure}[htbp]
\centering
\includegraphics[width=6.8cm]{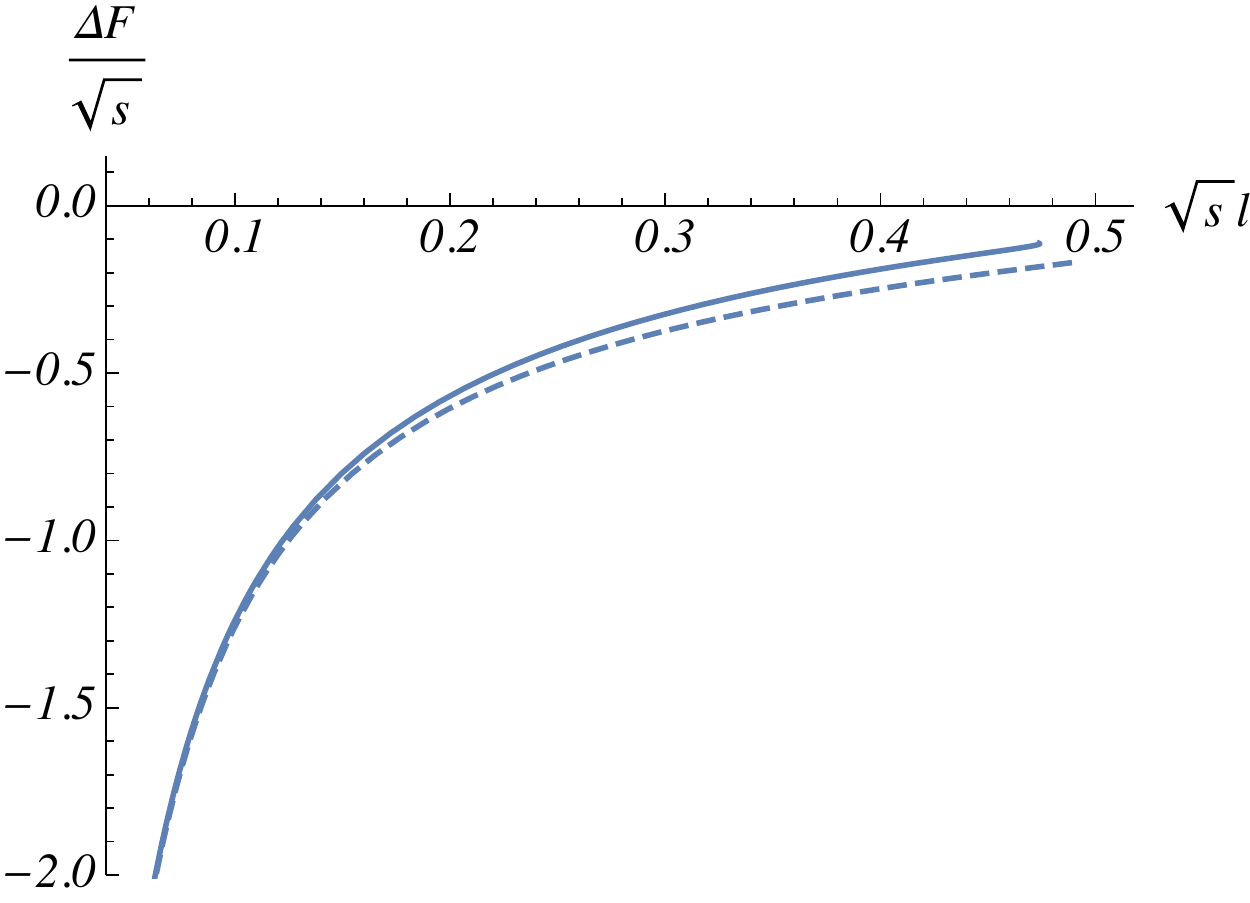}
\hspace{1.8cm}
\includegraphics[width=6.8cm]{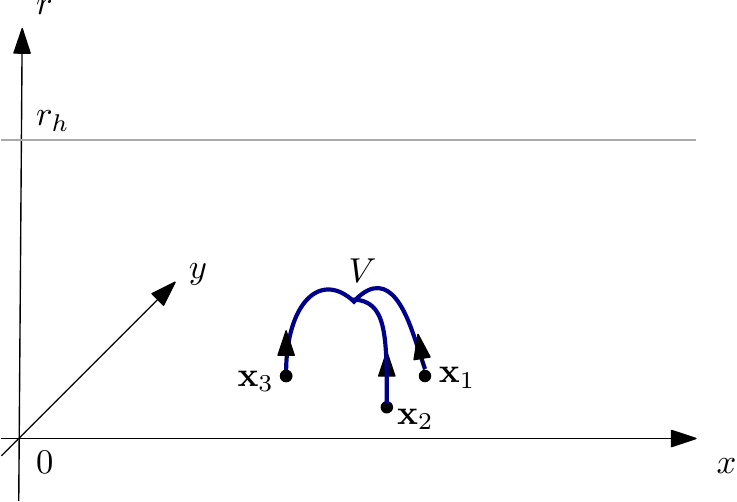}
\caption{{\small Left: Binding free energies of two-quark states at $T=0.345\sqrt{\s}$ and $\mu=0$. The solid curve corresponds to $\Delta F^{\bar 3}_{\qq}$ and the dashed one to $\oh \Delta F^{1}_{\qqb}$. We set $\g=0.176$, $\mathfrak{w}=-2.05$, and $\kappa=-0.102$. Right: A baryon configuration. $V$ is a baryon vertex. The quarks are placed at $\mathbf{x}_i$. For negative values of $\kappa$, gravity pulls the vertex toward the boundary that blunts the tip of the configuration.}
\\\vspace{.4cm}
}
\label{lipkin}
\end{figure}
Now a question arises: What happens at larger separations? In Figure \ref{lipkin} on the left, we display our results for the binding free energies. Here we set the values of the parameters so as to fit the model to the data for a pure $SU(3)$ gauge theory as close as possible. Clearly, for larger $l$ the binding free energy $\Delta F^{\bar 3}_{\qq}$ grows a bit faster than $\oh \Delta F^{1}_{\qqb}$. Thus our model
predicts 

\begin{equation}\label{1/2rule}
\Delta F^{\bar 3}_{\qq}\geq\oh \Delta F^1_{\qqb}
\,.
\end{equation}
The reason for this is the higher order terms in the expansion. For instance, the coefficient in front of the linear term in the expansion of $\Delta F^{\bar 3}_{\qq}$, which is an effective string tension inside a heavy diquark \cite{baryonsN}, turns out to be larger than what is expected from Lipkin's rule. One can think of \eqref{1/2rule} as a lower bound on $\Delta F^{\bar 3}_{\qq}$.

(ii) One might expect that things would be not so different for multi-loop correlators, except 
some technical points. We will not try to compute a generic correlator here. Rather, our goal is to briefly discuss new features related to three-quark interactions. So we want to consider a correlator of three similarly oriented loops.

The analysis proceeds in the following way. Let $\mathbf{x}_i$ be at the vertices of an equilateral triangle of length $l$. The string configuration which dominates at small $l$ is sketched in Figure \ref{lipkin}, on the right. If one normalizes the correlator as

\begin{equation}\label{C30-nor}
\mathbf{C}_{3,0}(l)=\frac{\langle\, L (\mathbf{x}_1) L(\mathbf{x}_2) L(\mathbf{x}_3)\,\rangle}{\langle\, L\,\rangle^3}
\,,
\end{equation}
then it determines a difference between the free energies: $\Delta F_{\3q}=F_{\3q}-3F_{\qu}$. The analysis of the correlator can be performed along the lines of \cite{baryonsN}. It shows that the short distance behavior of $\Delta F_{\3q}$ is 

\begin{equation}\label{F3q}
\Delta F_{\3q}(l)=-3\frac{\alpha_{\3q}}{l}+3F_0-T\ln w^{(1)}_{3,0} +o(l)
\,,
\end{equation}
with 
\begin{equation}\label{alpha3q}
\alpha_{\3q}=-\frac{\sqrt{3}}{4}\g\,{\cal B}\bigl(\kappa^2;\tfrac{1}{2},\tfrac{3}{4}\bigr)
\Bigl(\kappa\rho^{-\frac{1}{4}}+\frac{1}{4}{\cal B}\bigl(\kappa^2;\tfrac{1}{2},-\tfrac{1}{4}\bigr)\Bigr)
\,,
\qquad
\rho=1-\kappa^2
\,.
\end{equation}
Here ${\cal B}(z;a,b)=B(a,b)+B(z;a,b)$ and $B(a,b)$ is the beta function. $w^{(1)}_{3,0}$ is a weight of configuration. The coefficient $\alpha_{\3q}$ is the same as that in the static three-quark potential at zero temperature \cite{baryonsN}. 

The above formula has to be understood in a formal sense if $l$ is close to a critical value above which the baryon configuration of Figure \ref{lipkin} does not exist. We are in that situation here as in Sec.III. The bottom line is that we are unable to compute the exponential fall-off of correlators at long distances from the first principles. Once again, we follow the approach relying on a knowledge acquired from numerical simulations. Let us summarize what is needed for the purposes of the present paper. The three-quark system can be in a color singlet, two symmetrically mixed octet as well as in a color decuplet 
state. The three-quark free energy gets the contributions from all these states. At short distances, it is dominated by that of the singlet state. One of the parameterizations of the singlet free energy, is motivated by the $\Delta$-law which relates it to the binding free energy of the diquark system in the antitriplet channel \cite{hueb}. Explicitly

\begin{equation}\label{F3q-F2q}
\Delta F^1_{\3q}=3 \Delta F_{\qq}^{\bar 3}
\,.
\end{equation}
Like in the case of the there-quark potential at zero temperature, it should be a good starting point, especially at small and intermediate quark separations. 

Going back to our problem, one immediate conclusion is that the baryon configuration of Figure \ref{lipkin} should correspond to the singlet state. If so, then combining \eqref{F3q} with \eqref{F3q-F2q}
 and \eqref{F3} yields

\begin{equation}\label{m-bar}
m=\frac{1}{\alpha_{\3q}}\Bigl(F_0-\frac{1}{3}T\ln w^{(1)}_{3,0}\Bigr)
\,,
\end{equation}
which is one more estimate of the Debye mass.

\begin{figure}[htbp]
\centering
\includegraphics[width=6.8cm]{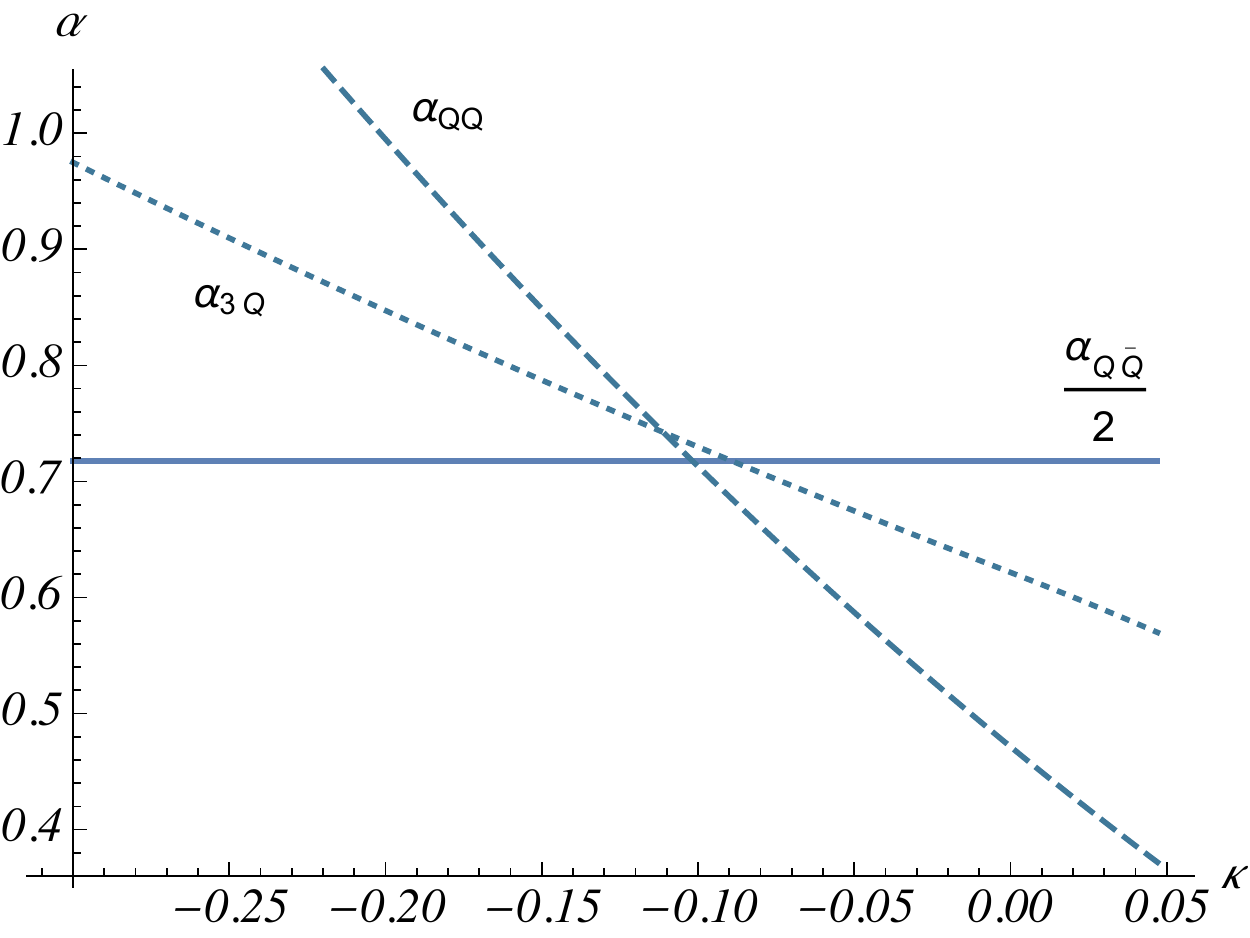}
\hspace{2cm}
\includegraphics[width=6.5cm]{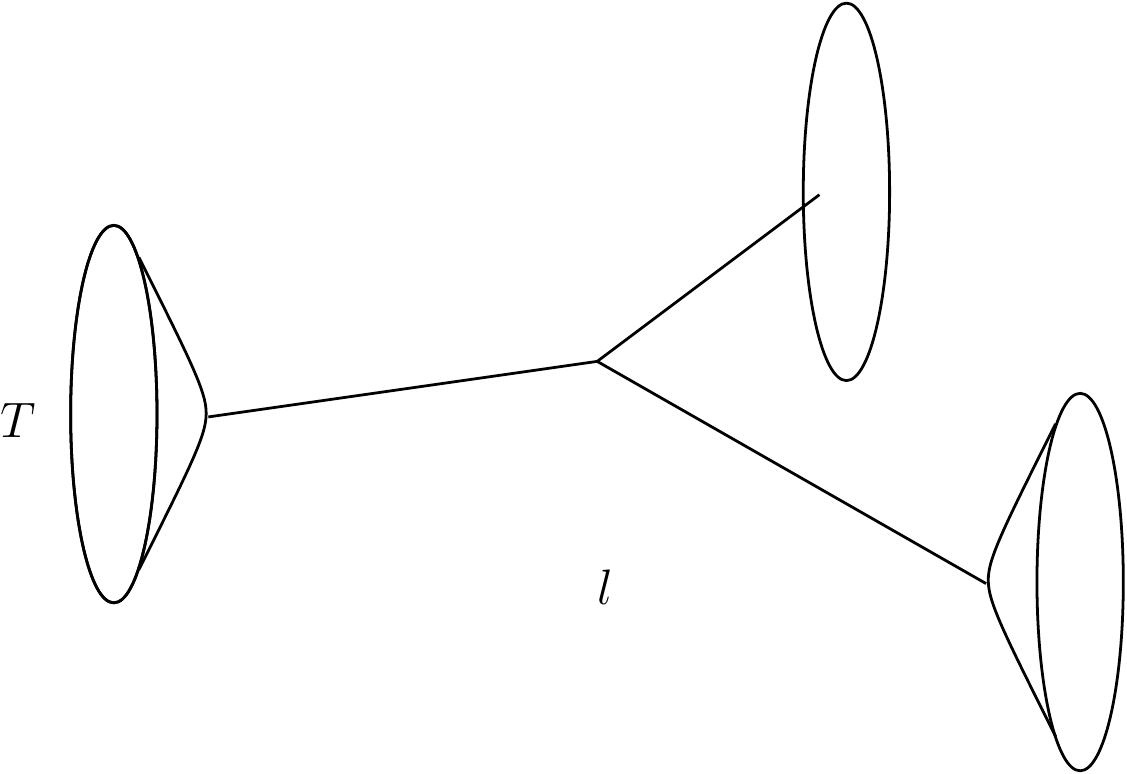}
\caption{{\small Left: $\oh\alpha_{\qqb}$, $\alpha_{\qq}$, and $\alpha_{\3q}$ in units of $\g$. The intersection points are at $\kappa\approx -0.111,\,-0.102,\,-0.089$. Right: Closed string states propagating between three boundary states placed at the vertices of an equalateral triangle.}
\\\vspace{.25cm}
}
\label{kappa}
\end{figure}

Since we have made the third estimate, it is time to ask about its consistency with 
the first two. At zero temperature, it is required $\oh\alpha_{\qqb}=\alpha_{\qq}=\alpha_{\3q}$. As seen from Figure \ref{kappa} on the left, this system of equations has no exact solutions.
However, if one sets $\alpha_{eff}=\oh(\alpha_{qq}(-0.111)+\oh\alpha_{\qqb})$, then its discrepancy from the others does not exceed $2\%$ that looks quite good in the light of effective theory. At finite temperature, it is required, in addition, that $\bigl(w^{(1)}_{1,\bar 1}\bigr)^\oh=w^{(1)}_{2,0}=\bigl(w^{(1)}_{3,0}\bigr)^{\frac{1}{3}}$. Like in Sec.III, we have no idea how to demonstrate, if possible, those relations. We can only appeal to lattice gauge theory, where the relative weights are $\frac{1}{9}$, $\frac{1}{3}$, and $\frac{1}{27}$, respectively.

The last point to be mentioned here is the large distance behavior of the correlator. It is convenient to express $\mathbf{C}_{3,0}$ in terms of the connected parts 

\begin{equation}\label{LLL-conn}
\mathbf{C}_{3,0}(l)=1+3\,\mathbf{C}^{\con}_{2,0}(l)+\mathbf{C}_{3,0}^{\con}(l)
\,,
\end{equation}
where $\mathbf{C}^{\con}_{2,0}=\mathbf{C}_{2,0}-1$.\footnote{Note that it decays for $l\rightarrow\infty$ as $\mathbf{C}^{\con}_{2,0}\sim\ep^{-Ml}$.} Since our technique doesn't allow us to compute the large $l$ behavior directly, we will assume that, like in the case of the two-point correlator, in the large $l$ limit $\mathbf{C}_{3,0}^{\con}$ can be described in terms of closed string states propagating in flat space, as sketched in Figure \ref{kappa} on the right. The solid lines here are propagators like that of Figure \ref{cylinder}. The leading asymptotics is given by the lightest closed string state. If a cubic interaction vertex is local, then it is  

\begin{equation}\label{LLL-M}
\mathbf{C}_{3,0}^{\con}(l) \sim \ep^{-\sqrt{3}Ml}
\,,
\end{equation}
as $l\rightarrow\infty$. Thus, $\mathbf{C}_{3,0}^{\con}$ is subleading to the first two terms in \eqref{LLL-conn}. 

At high temperatures, perturbation theory predicts that the prefactor in $\mathbf{C}_{3,0}^{\con}$ is of order $l^{-3}$. This suggests that $\mathbf{C}_{3,0}^{\con}\sim (\Delta F^1_{\3q})^3$, which when combined with \eqref{Mm} and \eqref{LLL-M} gives 

\begin{equation}\label{F3q-l}
\Delta F^1_{\3q}(l)\sim \ep^{-\frac{2}{\sqrt{3}}ml}
\,.
\end{equation}
Of course, the above arguments are heuristic. Note, however, that they are similar to those in the case of the two-loop correlator, where $\mathbf{C}^{\con}_{1,\bar 1}\sim(\Delta F^1_{\qqb})^2$ and, as a result, $M=2m$ \cite{lattice-karsch}.

What conclusions can one draw from the exponential decay law \eqref{F3q-l}? First, $\Delta F^1_{\3q}$ decays for $l\rightarrow\infty$ a bit faster than $\Delta F_{\qq}^{\bar 3}$. Indeed, one has $\frac{2}{\sqrt{3}}\approx 1.15>1$. If so, then the relation \eqref{F3q-F2q} is not valid at large quark separations. The situation here reminds us of one story: the three-quark potential at zero temperature and its fitting by the $\Delta$ and $Y$-laws, according to which the $Y$-law always holds at long distances.\footnote{It took some time to check this statement numerically. See, e.g., \cite{3q} and references therein.} Curiously, the ratio between the exponential decay constants is the same as that between the coefficients in front of the linear terms in the $\Delta$ and $Y$-laws. It is 
$\frac{2}{\sqrt{3}}=\frac{\sigma_{\Delta}}{\sigma_Y}$, with $\sigma_{\Delta}=\sqrt{3}\sigma$ and $\sigma_Y=\frac{3}{2}\sigma$. Second, one can think of $\Delta F^1_{\3q}$ as a complicated function whose asymptotic behavior is given by equations \eqref{F3q-F2q} and \eqref{F3q-l}. If so, then it follows that 

\begin{equation}\label{inequality}
\Delta F^1_{\3q}\geq 3\Delta F^{\bar 3}_{\qq}
\,.
\end{equation}

A word of caution here is that the model we are considering is somewhat crude to account 
for a small deviation even at short distances. Therefore, it would be very interesting to see the results of high-precision numerical studies of the quark free energies.\footnote{Unfortunately, the present precision \cite{hueb} does not allow one to make certain conclusions about the relations between the free energies.}

(iii) By now there are strong indications that the $SU(3)$ theory of quarks and gluons has a dual description in terms of quantum strings. Because the precise formulation of the latter is not known, we can only gain useful insight and grow with each success of the effective string model already at our disposal. Despite its efficiency \cite{az1,a-pol,baryonsN}, one might think of criticizing this model on several grounds including, (a) use of ad hoc background geometry whose consistency remains an open question within the $\alpha'$-expansion of string theory, (b) inability to account for corrections arising from string fluctuations to the classical expression \eqref{C}, (c) use of string theory to describe QCD with a finite number of colors and flavors, and (d) a lack of a satisfactory framework for string perturbation theory in the presence of baryon (antibaryon) vertices. We have nothing to say at this point, other than that we hope to return to these issues in future work.
\begin{acknowledgments}
We are grateful to C. Bachas, S. Finazzo, Y. Maezawa, J. Noronha, and P. Weisz for helpful discussions and correspondence. We also thank the Arnold Sommerfeld Center for Theoretical Physics for the warm hospitality. This work was supported by Russian Science Foundation grant 16-12-10151. 
\end{acknowledgments}
\appendix
\section{A static Nambu-Goto string with fixed endpoints}
\renewcommand{\theequation}{A.\arabic{equation}}
\setcounter{equation}{0}

The goal here is to generalize the results of \cite{baryonsN} to the background fields of \eqref{metric5}. This will enable us to build some of the string configurations in concrete examples. We set $\rh<1/\sqrt{\s}$ because it guarantees that the corresponding gauge theory is deconfined \cite{az2}.

As usual, the Nambu-Goto action is given by

\begin{equation}\label{NG}
S_{NG}=\frac{1}{2\pi\alpha'}\int_0^1 d\sigma\int d\tau\,\sqrt{\gamma}
\,,
\end{equation}
with $\gamma$ an induced metric on a world-sheet with Euclidean signature. However, this is not the whole story. Since the quarks and antiquarks are regarded as string endpoints\footnote{In our approach free strings are associated with mesons.} and the $U(1)$ background field is introduced to mimic the baryonic charge in gauge theories, there are additional coupling terms to the Nambu-Goto action 

\begin{equation}\label{SA}
S_b=\frac{1}{3}\int d\tau A_t \frac{d t}{d \tau}\,\biggl\vert_{\sigma=0}
-\frac{1}{3}\int d\tau A_t \frac{d t}{d \tau}\,\biggr\vert_{\sigma=1}
\,.
\end{equation}
In other words, the string endpoints are electrically charged with respect to the background gauge field, with a charge that is $+1/3$ for a quark and $-1/3$ for an antiquark. Note that our normalization of the baryonic charge is chosen so that a baryon has a charge of $+1$.

Without of loss of generality, assume that a string is stretched between the two fixed points $P$ and $B$ in the $xr$-plane,  
\begin{figure}[htbp]
\includegraphics[width=5.2cm]{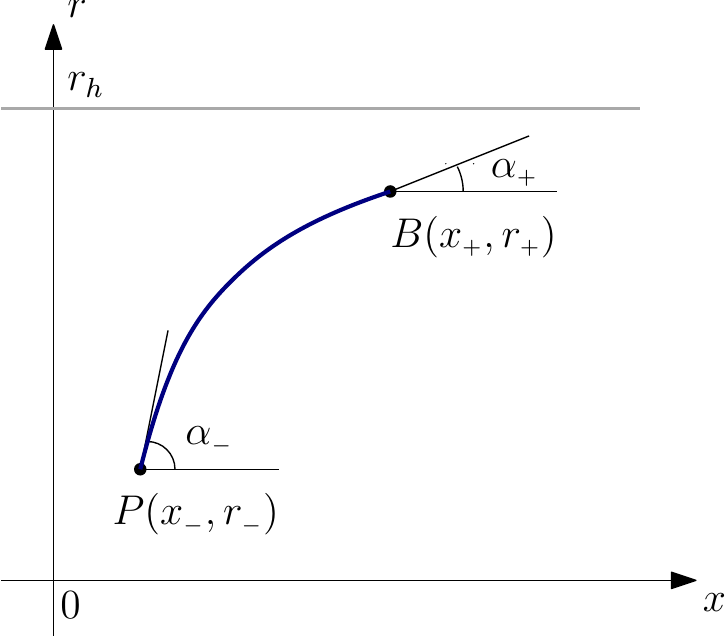}
\caption{{\small A string stretched between a quark placed at $P$ and an antiquark placed at $B$. $\alpha_{\scriptscriptstyle\pm}$ denote the tangent angles, which are both assumed to be positive. The horizon is at $r=\rh$, while the boundary is at $r=0$.}}
\label{ng}
\end{figure}
\noindent as shown in Figure \ref{ng}. This means that at the endpoints we have

\begin{equation}\label{string-bc}
x(0)=\xm\,,\quad x(1)=\xp\,,\quad r(0)=\rm\,,\quad r(1)=\rp
\,.
\end{equation}

Since we are interested in a static configuration, we choose the static gauge $t=\tau$ and $x=a\sigma +b$, where $a=\xp-\xm$ and $b=\xm$. Moreover, in \eqref{NG} and \eqref{SA} the integrands are time-independent, so the integration over $t$ simply gives a factor of $1/T$. We then have

\begin{equation}\label{total-action}
S=\frac{1}{T}
\biggl(\g\int_{\xm}^{\xp} dx\,w(r)\sqrt{f(r)+(\partial_x r)^2}\,\,+\frac{1}{3}A_t(\rm)-\frac{1}{3}
A_t(\rp)\biggr)
\,.
\end{equation}
For convenience, we use the shorthand notation $\g=\frac{{\cal R}^2}{2\pi\alpha'}$ and $\partial_x r=\frac{\partial r}{\partial x}$. 

Obviously, the variation of the action \eqref{total-action} includes the contributions from the endpoints. In our study, we are interested in the configurations shown in Figures \ref{conf1-1} and \ref{conf2}. In all those cases, there are no such contributions if the field $r$ is subject to the Dirichlet boundary condition on the boundary of space and the Neumann boundary condition at string junctions. The latter deserves a separate discussion. We will return to this issue after analyzing the equation of motion following from the above action.

Since the integrand in \eqref{total-action} does not depend explicitly on $x$, the corresponding Euler-Lagrange equation has the first integral

\begin{equation}\label{I}
I=\frac{w f}{\sqrt{f+(\partial_x r)^2}}
\,,
\end{equation}
which takes the form 
\begin{equation}\label{I-PB}
I=\frac{wf}{\sqrt{f+\tan^2\alpha_{\scriptscriptstyle\pm}}}\,\biggl\vert_{r=r_{\scriptscriptstyle \pm}}
\,
\end{equation}
at the endpoints, where $\tan\alpha_{\scriptscriptstyle \pm}=\partial_xr\vert_{x=x_{\scriptscriptstyle \pm}}$.

In general, $\ap$ can take both positive or negative values. For our purposes, we need only positive ones such that a function $r(x)$ describing a string profile is increasing on the interval $[\xm,\xp]$, as sketched in Figure \ref{ng}. In this case, \eqref{I} yields a differential equation $\partial_x r=\bigl(f^2\frac{w^2}{I^2}-f\bigr)^\oh$ that can be easily integrated over the variables $x$ and $r$. So, we get

\begin{equation}\label{l0}
\ell=\xp-\xm=\int^{\rp}_{\rm} \frac{dr}{\sqrt{f}}\, 
\bigl(\eta^{-1}-1\bigr)^{-\frac{1}{2}}
\,,\qquad\text{with}\qquad \eta=\frac{I^2}{f w^2}
\,.
\end{equation}
After a rescaling of $r$, the integral becomes
\begin{equation}\label{l}
\ell=\rp\int^{1}_{\frac{\rm}{\rp}} \frac{dv}{\sqrt{f}}\, 
\bigl(\eta^{-1}-1\bigr)^{-\frac{1}{2}}
\,.
\end{equation}

Having found the solution, we can now compute the corresponding action. It includes the minimal area and boundary contributions. The result is 

\begin{equation}\label{S}
S=\frac{1}{T}
\biggl(
\g\rp\int^{1}_{\frac{\rm}{\rp}} dv\,w 
\bigl(1-\eta\bigr)^{-\frac{1}{2}}
\,\,
+\frac{1}{3}A_t(\rm)-\frac{1}{3}A_t(\rp)\biggr)
\,.
\end{equation}
In the case of a string stretched along the $r$-direction, the integral can be evaluated analytically, yielding

\begin{equation}\label{S-vert}
S=\frac{1}{T}
\biggl(
\g\biggl[
\sqrt{\pi\s}
\Bigl(\erfi(\sqrt{\s}\rp)-\erfi(\sqrt{\s}\rm)\Bigr)+
\frac{\ep^{\s\rm^2}}{\rm}-\frac{\ep^{\s\rp^2}}{\rp}
\biggr]
\,\,
+\frac{1}{3}A_t(\rm)-\frac{1}{3}A_t(\rp)
\biggr)
\,.
\end{equation}

Another special case is a string with $P$ on the boundary of space, i.e. when $\rm=0$. In this case, the integral \eqref{l} remains finite at $v=0$, while \eqref{S} tends to infinity. We regularize it by imposing a cutoff $\epsilon$ such that $r\geq \epsilon$. In the limit $\epsilon\rightarrow 0$, the regularized expression behaves like 

\begin{equation}
S_R=\frac{\g}{T\epsilon}+S+O(\epsilon)
\,.
\end{equation}
Subtracting the $\frac{1}{\epsilon}$ term and letting $\epsilon=0$, we get a finite result 

\begin{equation}\label{S0}
S=\frac{1}{T}
\biggl(
\g\rp\int^{1}_{0} dv\,\biggl[w 
\bigl(1-\eta\bigr)^{-\frac{1}{2}}
-\frac{1}{\rp^2v^2}-\frac{1}{\rp^2}\biggr]
\,\,
+\mu_q-\frac{1}{3}A_t(\rp)\biggr)
+c
\,,
\end{equation}
where $\mu_q$ is a quark chemical potential and $c$ is a normalization constant. It is dimensionless and related to the one usually used in the literature by a simple rescaling. In the last step, the boundary condition satisfied by $A_t(r)$ at $r=0$ and relation $\mu=3\mu_q$ were used.

We similarly deal with the expression \eqref{S-vert}. As a result, we obtain

\begin{equation}\label{S0-vert}
S=\frac{1}{T}
\biggl(
\g\biggl[
\sqrt{\pi\s}\erfi(\sqrt{s}\rp)
-
\frac{\ep^{\s\rp^2}}{\rp}
\biggr]
\,\,
+\mu_q-\frac{1}{3}A_t(\rp)
\biggr)+c
\,.
\end{equation}
For $\rp=\rh$, when a string stretched between the boundary and the horizon, this expression simplifies to

\begin{equation}\label{S-pol}
S=\frac{1}{T}
\Bigl(\mu_q-F_0 \Bigr)+c
\,,
\qquad\text{with}
\qquad
F_0=\g\biggl[\frac{\ep^{\s\rh^2}}{\rh}-\sqrt{\pi\s}\erfi(\sqrt{\s}\rh)\biggr]
\,.
\end{equation}


\section{Gluing conditions}
\renewcommand{\theequation}{B.\arabic{equation}}
\setcounter{equation}{0}

\begin{figure}[htbp]
\includegraphics[width=5.5cm]{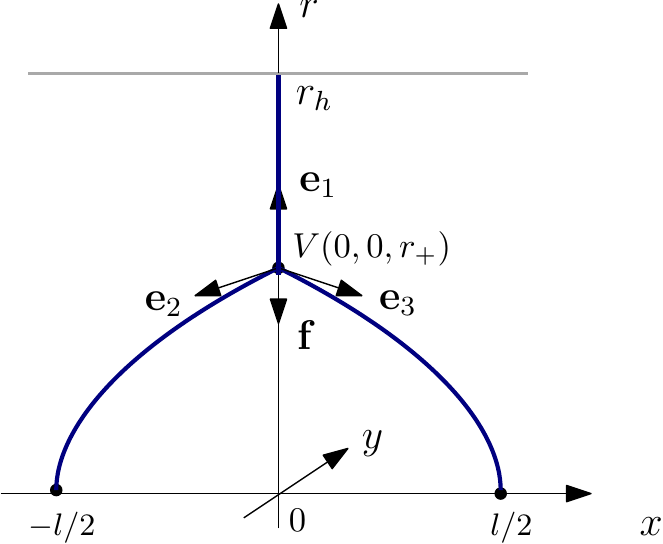}
\caption{{\small Three strings meeting at a string junction (vertex) placed at $V$. The black hole horizon is at $r=\rh$.
}}
\label{fb}
\end{figure}

The string solutions we discussed in the Appendix A provide just the basic blocks for building multi-string configurations. By analogy with what was done in \cite{baryonsN}, we need certain gluing conditions for such blocks in the presence of the background fields \eqref{metric5}. Here we will illustrate the idea with a simple example, as in Figure \ref{fb}, that suffices for 
computing the free energy of a quark pair (diquark). 

A physical way of interpreting the gluing conditions is to say that a static string configuration must obey the condition that a net force vanishes at any vertex. We then 
write \footnote{We denote vectors by boldface letters.}

\begin{equation}\label{v-netforce}
\sum_{i=1}^3\mathbf{e}_{i}+\mathbf{f}=0
\,,
\end{equation}
where $\mathbf{e}_i$ is a tangent vector at a string endpoint which represents a force exerted by the $i$ string on the vertex and $\mathbf{f}$ is a gravitational force exerted on the vertex.

In the presence of the $U(1)$ gauge field, the zero force condition is to be completed by the neutrality condition

\begin{equation}\label{v-netcharge}
\sum_{i=1}^3q_{i}+Q=0
\,,
\end{equation}
where $q_i$ is a charge of an endpoint of the $i$ string that ends on the vertex. It 
is $+1$ or $-1$ depending on the particle type (quark or antiquark). $Q$ is a charge of the vertex that is $+3$ or $-3$ depending on the type of the vertex in question (baryon or antibaryon). 

For a mathematical treatment of the configuration of Figure \ref{fb}, we consider the total action of three strings and a vertex in the presence of the $U(1)$ background gauge field. The neutrality condition \eqref{v-netcharge} implies that the terms linear in $A_t(\rp)$ cancel each other. This simplifies the problem of finding the variation of the total action with respect to a location of the vertex and associated zero force condition. The former is nothing else but the Neumann boundary condition at $r=\rp$. 

With this simplification, we can straightforwardly extend the analysis of \cite{baryonsN} to the black hole geometry \eqref{metric5}. Accordingly, we take 

\begin{equation}\label{NGaction}
S=\sum_{i=1}^3 S_i\,+S_{\text{vert}}
\,,
\end{equation}
where $S_i$ is the Nambu-Goto action of the $i$-string and $S_{\text{vert}}$ is that of the vertex. 

Since we are interested in the static configuration, it is convenient to choose gauge conditions 

\begin{equation}\label{gauges}
t_i(\tau_i)=\tau_i\,,
\qquad
x_i(\sigma_i)=a_i\sigma_i+b_i
\,,
\end{equation}
with $(\tau_i,\sigma_i)$ the world-sheet coordinates. Then the action of the $i$-string takes the form 

\begin{equation}\label{NGaction2}
S_i=\frac{\g}{T}\int_0^1 d\sigma_i w \sqrt{f a_i^2+\bigl(\partial_{\sigma_i} r\bigr)^2}
\,,
\end{equation}
where a partial derivative with respect to $\sigma_i$ is conveniently denoted $\partial_{\sigma_i}$.
 
For the configuration in question, the boundary conditions on the fields are given by 

\begin{equation}
x_1(0)=x_i(1)=0\,,\quad
x_2(0)=-x_3(0)=-\ell/2\,,\quad
r_1(0)=\rh\,,\quad
r_2(0)=r_3(0)=0\,,\quad
r_i(1)=\rp\,.
\end{equation}
With these boundary conditions, a short calculation shows that $a_1=b_1=0$ and $a_2=-b_2=-a_3=b_3=\ell/2$. 

Since the configuration has a reflection symmetry, it lies entirely in the $xr$-plane and, as a consequence, the equations for the $x$ and $y$ components of the net force are trivial. Thus, we need only consider the variation with respect to $\rp$ keeping in mind that $\delta r_i(1)=\delta\rp$. After some straightforward computation, we obtain 

\begin{equation}\label{balance}
\frac{2\tan\ap}{\sqrt{f+\tan^2\ap}}-1+3\kappa\rp^2\ep^{-\s\rp^2}\partial_{\rp}
\Bigl(\frac{\ep^{-2\s\rp^2}}{\rp}\sqrt{f}\,\Bigr)=0
\,,
\end{equation}
where $\kappa=\frac{m}{3\g}$ and $\ap=\ap{}_2=-\ap{}_3$. The last term results from the gravitational force exerted on the vertex.

Importantly, the resulting equation allows us to express $\ap$ in terms of $\rp$. We have 

\begin{equation}\label{tan}
\tan^2\ap=\frac{f\,\theta^2}{4-\theta^2}
\,,
\qquad\text{with}
\qquad
 \theta=1-3\kappa\rp^2\ep^{-\s\rp^2}\partial_{\rp}
 \Bigl(\frac{\ep^{-2\s\rp^2}}{\rp}\sqrt{f}\,\Bigr)
 \,.
\end{equation}


\section{Consistency Issues}
\renewcommand{\theequation}{C.\arabic{equation}}
\setcounter{equation}{0}

It is well known that Weyl invariance is essential to the consistency of string theory \cite{joe}. Apart from being a necessary requirement for a two-dimensional field theory on a worldsheet, this is very attractive for phenomenology as it allows one to reduce a number of free parameters. As an important illustration of some of these ideas, we will discuss the relation between $w$, $f$, and $A_t$ in \eqref{metric5}.
 
For our purposes in this paper, what we need to know is that the Weyl coefficient for $f$ is essentially the renormalization group beta function $\beta^f$ \cite{joe}. Given the beta function for the metric $\beta^G_{\mu\nu}=\Lambda\frac{dG_{\mu\nu}}{d\Lambda}$, with $\Lambda$ a worldsheet scale, 
the beta function for $f$ is determined by the components $G_{tt}$ and $G_{ii}$ which respectively represent the time and spatial (any one) components of the metric. The formula reads
 
\begin{equation}\label{beta-f}
\beta^f=\frac{1}{w{\cal R}^2 }\Bigl(\beta^G_{tt}-f\beta^G_{ii}\Bigr)
\,.
\end{equation}
Here $i$ takes any value from $1$ to $3$.

As a warmup, consider the case $A_t=0$. In the context of the $\alpha'$ expansion of $\beta^G_{\mu\nu}$, the leading term is given by the Ricci tensor \cite{joe} and one gets 

\begin{equation}\label{beta-f2}
\beta^f=
\frac{\alpha'}{w{\cal R}^2}\Bigl(R_{tt}-f R_{ii}\Bigr)
\,.
\end{equation}
By letting $\beta^f=0$ and then $R_{tt}-f R_{ii}=0$, one further deduces

\begin{equation}\label{kiri}
\partial_r^2f+\frac{3}{2}\frac{\partial_r w}{w}\partial_r f=0
\,,
\end{equation}
with $\partial_r=\frac{\partial}{\partial r}$. This provides the desired relation between $w$ and $f$. Such a relation occurs in models of Einstein gravity coupled to matter\footnote{In the case of dilaton gravity, it was discussed in detail in \cite{kiritsis}.}, if the matter energy-momentum tensor obeys the constraint

\begin{equation}\label{T}
T_{tt}=f\,T_{ii}
\,.
\end{equation}
Then $R_{tt}-f R_{ii}=0$ is a consequence of the Einstein equations.

When we add the $U(1)$ gauge field, we gain a Weyl coefficient $\beta^A_{\mu}$. To leading order in $\alpha'$, it is given by 

\begin{equation}\label{beta-A}
\beta_\mu^A=\alpha'\nabla^\nu F_{\mu\nu}
\,.
\end{equation}
For the background fields \eqref{metric5}, we have, by letting $\beta^A_\mu=0$,

\begin{equation}\label{ddA}
\partial_r\bigl(w^{\frac{1}{2}} \partial_r A_t\bigr)=0
\,.
\end{equation}
This equation with the boundary conditions \eqref{A-boundary} can immediately be integrated with the result\footnote{The difference with \cite{a-mu} stems from the fact that we are dealing with the $5d$ background metric. Also, the imaginary unit $i$ is excluded from the solution.}

\begin{equation}\label{A0}
A_t(r)=\mu-\frac{c_0}{{\cal R}\s}\Bigl(1-\ep^{-\oh\s r^2}\Bigr)
\,,
\end{equation}
where $c_0$ is a positive constant. This form of the solution is particularly convenient as it allows one to make a contact with the Reissner-Nordstr\"om solution of \cite{chamblin} in the limit $\s\rightarrow 0$.

Another gain from the gauge field is that the Weyl coefficient $\beta^G_{\mu\nu}$ receives $A$-dependent corrections. To leading order in $\alpha'$, there are contributions from two terms: $\alpha'^2 G^{\lambda\sigma}F_{\mu\lambda}F_{\nu\sigma}$ and $\alpha'^2 G_{\mu\nu}F^2$. The latter is clearly irrelevant for our purposes, but the former matters because it has impact on $\beta^f$. The contribution coming from the coefficient $\beta^G_{tt}$ is $c_1\tfrac{\alpha'^2f}{{\cal R}^2w}(\partial_rA_t)^2$.\footnote{Here we are assuming a numerical factor $c_1$ which can be traced into a ratio between the gauge and gravitational couplings in the effective Lagrangian $R+\Lambda+g F^2$ of \cite{chamblin} or, explicitly $c_1=\frac{4\pi\g}{{\cal R}^2}g$. See also \cite{colangelo2}.} Then equation \eqref{beta-f2} becomes

\begin{equation}\label{beta-fA}
\beta^f=
\frac{\alpha'}{w{\cal R}^2}\biggl(R_{tt}-f R_{ii}+\frac{c_1}{2\pi\g}fw^{-1}\bigl(\partial_r A_t\bigr)^2
\biggr)
\,.
\end{equation}
Here we used the fact that $\alpha'={\cal R}^2/2\pi\g$. By letting $\beta^f=0$, we arrive at the desired equation

\begin{equation}\label{an}
\partial_r^2f+\frac{3}{2}\frac{\partial_r w}{w}\partial_r f-\frac{c_0^2c_1}{\pi\g{\cal R}^2}w^{-2}=0\,.
\end{equation}
Like in the previous case, it can be integrated. Taking into account the boundary conditions \eqref{G-boundary}, we obtain

\begin{equation}\label{fA}
f(r)=1-A\Bigl(1-\bigl(1+2\s r^2\bigr)\ep^{-2\s r^2}\Bigr)-B
\Bigl(1-\bigl(1+\tfrac{3}{2}\s r^2\bigr)\ep^{-\tfrac{3}{2}\s r^2}\Bigr)
\,.
\end{equation}
Here $A=\frac{c_0^2c_1}{8\pi\g{\cal R}^2\s^3}$ and $B=\frac{1-A(1-(1+2h)\ep^{-2h})}{1-(1+\frac{3}{2}h)\ep^{-\frac{3}{2}h}}$, with $h=\s\rh^2$. Note that at $c_0=0$ the solution reduces to that of \cite{noro}. 

With the expressions for $A_t$ and $f$, the temperature and baryon chemical potential are defined as usual

\begin{equation}\label{T-mu}
T=\frac{1}{4\pi}\Big\vert\partial_r f\Big\vert_{r=\rh}
\,,\qquad
\mu=A_t(0)
\,.
\end{equation}
We think of $c_0$ and $h$ as two variables and of $c_1$ as a parameter. However, it is convenient for practical purposes to introduce a new variable $q^2=\frac{A}{36}(9+(7+6h)\ep^{-2h}-16\ep^{-\frac{1}{2}h})$ and to express $T$ and $\mu$ in terms of $q$ and $h$:

\begin{equation}\label{tmu}
\frac{T}{\sqrt{\s}}=\frac{9}{8\pi}\,\frac{(1-q^2)\,h^{\frac{3}{2}}}{
U(h)}
\,,\qquad
\frac{\mu}{\sqrt{\s}}=2\sqrt{3}\mathfrak{r}\,\frac{q}{Z^{\oh}(h)}
\,,
\end{equation}
with 
\begin{equation}\label{UZ}
\mathfrak{r}=\sqrt{\frac{6\pi\g}{c_1}}\,,
\qquad
U(h)=\ep^{\frac{3}{2}h}-1-\tfrac{3}{2}h
\,,
\qquad
Z(h)=\frac{9+(7+6h)\ep^{-2h}-16\ep^{-\oh h}}{1-2\,\ep^{-\oh h}+\ep^{-h}}
\,.
\end{equation}
In this form, $c_1$ combines with $\g$ to form a new parameter $\mathfrak{r}$ to which we can assign arbitrary values. It is easy to guess the domain of the functions $T$ and $\mu$. It is defined by inequalities $0\leq q\leq 1$ and $0<h<1$. In particular, if $q=0$, then $\mu=0$ and if $q=1$, then $T=0$. 

Finally, it is worth noting that the behavior of $T$ and $\mu$ near $h=0$ is 

\begin{equation}\label{small-h}
\frac{T}{\sqrt{\s}}=\frac{1-q^2}{\sqrt{h}}
\,,
\qquad
\frac{\mu}{\sqrt{s}}=\frac{\mathfrak{r}q}{\sqrt{h}}
\,.
\end{equation}


\section{Taylor Series of $M$}
\renewcommand{\theequation}{D.\arabic{equation}}
\setcounter{equation}{0}

Our goal here is to compute the first few coefficients of the Taylor series of $M(T,\mu)$ about the points  $(T,0)$ and $(0,\mu)$. The higher order coefficients can be computed in a similar fashion.

Before getting started with the Taylor series of $M$, let us rewrite the relations \eqref{tmu} in a more convenient form 

\begin{equation}\label{h}
8xU=9
h^{\frac{3}{2}}\bigl(1-Zy^2\bigr)
\,,
\end{equation}
where $x=\pi\tfrac{T}{\sqrt{\s}}$ and $y=\frac{1}{2\sqrt{3}\mathfrak{r}}\frac{\mu}{\sqrt{\s}}$. It now remains to obtain $h$ as a function of $x$ and $y$. We are unable to do so analytically because the above equation is highly nonlinear. Therefore we use the Taylor expansion of $h$ that suffices for our present needs.

We start by illustrating the procedure in the case of small $y$. The form of \eqref{h} implies that $h$ is an even function of $y$. So we expand $h$ as 

\begin{equation}\label{h-exp}
h(x,y)=h_0(x)+h_2(x)y^2+O(y^4)
\,.
\end{equation}
Inserting \eqref{h-exp} into \eqref{h} and equating the terms with identical powers of $y$ on the two sides of equation gives 

\begin{equation}\label{h0h2}
x=\frac{9}{8}\h0^{\frac{3}{2}}U^{-1}
\,,
\qquad
h_2=\frac{2}{3}
\frac{\h0 ZU}
{U+\h0(1-\ep^{\frac{3}{2}\h0})}
\,\,.
\end{equation}
The first equation says that $x$ is a function of $h_0$, while the second equation represents a recursion relation for $h_2$ once $h_0$ is known. Here again, we are unable to analytically find $h_0$ as a function of $x$. 

We are now ready to compute the two leading coefficients of the Taylor series \eqref{mu-expansion}. Taking the solution for $h$ and using \eqref{M}, after a short computation we obtain

\begin{equation}\label{M2}
M_0(\h0)=\frac{\gamma}{\pi}\sqrt{\s}
\biggl(\h0^{-\oh}\ep^{\h0}
-
\sqrt{\pi}\,\erfi\bigl(\sqrt{\h0}\,\bigr)
+\frac{9\mathfrak{w}}{8\pi}\h0^{\frac{3}{2}}U^{-1}
\biggr)
\,,
\qquad
M_2(\h0)=\frac{9\gamma\sqrt{\s}}{256\pi^3\mathfrak{r}^2}
\,
\frac{\h0^{\frac{5}{2}}\ep^{\h0}ZU^{-1}}{\h0(\ep^{\frac{3}{2}\h0}-1)-U}
\,.
\end{equation}
Thus, the Taylor coefficients are given by the parametric equations \eqref{h0h2} and \eqref{M2}, with $h_0$ a parameter.
 
It is instructive to see what happens to these coefficients as $h_0$ goes to zero. This limit is of interest because it corresponds to the high temperature limit, as follows 
from \eqref{small-h}. For the ratio between $M_0$ and $M_2$, we get 

\begin{equation}\label{M2/M0}
\frac{M_0}{M_2}=\pi\mathfrak{r}^2\bigl(\mathfrak{w}+\pi\bigr)
\,,
\end{equation}
which is simply a constant. What is important for us is that \eqref{M2/M0} shows a similar behavior to that known from perturbative QCD.

We can now carry out a precisely analogous computation for the case of small $x$. In that case, we expand the function $h$ as usual

\begin{equation}\label{h-expT}
h(x,y)=h_0(y)+h_1(y)x+O(x^2)
\,.
\end{equation}
After inserting it into equation \eqref{h} and equating the terms with identical powers of $x$, we find 

\begin{equation}\label{h0h1}
y=Z^{-\oh}(\h0)
\,,
\qquad
h_1(y)=\frac{8}{9}
\frac{\h0^{-\frac{3}{2}}(\ep^{\oh\h0}-1)^2UZ}{(\ep^{\oh\h0}-1)Z-8\ep^{-\h0}U}
\,.
\end{equation}
The first equation gives $y$ as a function of $\h0$, while the second gives a recursion relation for $h_1$. 

Given the solution \eqref{h0h1}, we can compute the two leading coefficients of the Taylor series \eqref{T-expansion} by using the same formula as before, with the result

\begin{equation}\label{M-T}
M_0(\h0)=\frac{\gamma}{\pi}\sqrt{\s}
\biggl(\h0^{-\oh}\ep^{\h0}
-
\sqrt{\pi}\,\erfi\bigl(\sqrt{\h0}\,\bigr)\biggr)
\,,\qquad
M_1(\h0)=2\sqrt{3}\gamma
\frac{\mathfrak{r}\sqrt{\s}}{Z^\oh}
\biggl(\frac{\mathfrak{w}}{\pi}+
\frac{4}{9}\,
\frac{\h0^{-3}\ep^{\h0}(\ep^{\oh\h0}-1)^2UZ}{8\ep^{-\h0}U-(\ep^{\oh\h0}-1)Z}
\biggr)
\,.
\end{equation}
Thus, the Taylor coefficients are given by the parametric equations \eqref{h0h1} and \eqref{M-T}, with $\h0$ is a parameter.

For completeness, we note that in the limit $\h0\rightarrow 0$, now corresponding to large chemical potential, the ratio between $M_1$ and $M_0$ tends to a constant value. Explicitly

\begin{equation}\label{M1/M0}
\frac{M_1}{M_0}=\mathfrak{r}\Bigl(\mathfrak{w}+\oh\pi
\Bigr)
\,.
\end{equation}


\end{document}